\documentclass[aps,pra,reprint,superscriptaddress,twocolumn]{revtex4-1}
\usepackage{amssymb}
\usepackage{amsmath}
\usepackage{graphicx}
\usepackage{epstopdf}
\usepackage{algorithm}
\usepackage{algpseudocode}
\algnewcommand\algorithmicinput{\textbf{Input:}}
\algnewcommand\INPUT{\item[\algorithmicinput]}
\algnewcommand\Break{\textbf{break}}
\begin{document}
\title{High Fidelity Entangling Gates in a 3D Ion Crystal under Micromotion}
\author{Y.-K. Wu}
\thanks{These authors contributed equally}
\affiliation{Center for Quantum Information, IIIS, Tsinghua University, Beijing 100084, P. R. China}
\author{Z.-D. Liu}
\thanks{These authors contributed equally}
\affiliation{Center for Quantum Information, IIIS, Tsinghua University, Beijing 100084, P. R. China}
\author{W.-D. Zhao}
\affiliation{Center for Quantum Information, IIIS, Tsinghua University, Beijing 100084, P. R. China}
\author{L.-M. Duan}
\affiliation{Center for Quantum Information, IIIS, Tsinghua University, Beijing 100084, P. R. China}
\date{\today}

\begin{abstract}
Ion trap is one of the most promising candidates for quantum computing. Current schemes mainly focus on a linear chain of up to about one hundred ions in a Paul trap. To further scale up the qubit number, one possible direction is to use 2D or 3D ion crystals (Wigner crystals). In these systems, ions are generally subjected to large micromotion due to the strong fast-oscillating electric field, which can significantly influence the performance of entangling gates. In this work, we develop an efficient numerical method to design high-fidelity entangling gates in a general 3D ion crystal. We present numerical algorithms to solve the equilibrium configuration of the ions and their collective normal modes. We then give a mathematical description of the micromotion and use it to generalize the gate scheme for linear ion chains into a general 3D crystal. The involved time integral of highly oscillatory functions is expanded into a fast-converging series for accurate and efficient evaluation and optimization. As a numerical example, we show a high-fidelity entangling gate design between two ions in a 100-ion crystal, with a theoretical fidelity above 99.9\%.
\end{abstract}

\maketitle

\section{Introduction}
Over the past few decades, quantum computing has attracted wide interest because it challenges the strong Church-Turing thesis and has a potential exponential speedup over any classical computers for certain problems \cite{nielsen2000quantum}.
To realize quantum computing, some basic requirements, known as the DiVincenzo's criteria \cite{divincenzo2000criteria}, need to be fulfilled, among which is a universal gate set that any desired multi-qubit unitary gates can be generated from.
A commonly used universal gate set consists of a few single-qubit gates and a two-qubit entangling gate \cite{nielsen2000quantum}. While the single-qubit gates are relatively simple, the two-qubit gate usually turns out to be much more difficult and has become the focus of research. Besides, such entangling gates also find applications in other fields such as digital quantum simulation \cite{RevModPhys.86.153}, variational hybrid quantum-classical algorithms \cite{farhi2014quantum,McClean_2016} and quantum metrology \cite{Giovannetti1330}.

Trapped ions have become one of the leading platforms for realizing quantum computing owing to the long coherence time, convenient initialization and readout, and the strong state-dependent coupling between ions mediated by laser  \cite{Cirac1995,RevModPhys.75.281,Blatt2008,Monroe2013} or microwave \cite{PhysRevLett.87.257904,PhysRevLett.108.220502,PhysRevLett.101.090502,ospelkaus2011microwave} driving. Dedicated schemes for high-fidelity entangling gates, known as the Molmer-Sorensen (MS) gate \cite{Sorensen1999} and its variants, have been developed for ion chains in a linear Paul trap. The gate can be designed using different collective motional modes of the ions \cite{PhysRevA.71.062309,PhysRevLett.97.050505}, through various control and optimization methods for the driving field \cite{Zhu2006,PhysRevLett.114.120502,PhysRevA.98.032318}, and have been demonstrated in experiments from two to more than ten ions \cite{d.wineland2000ion.ghz,leibfried2003experimental,PhysRevLett.106.130506,Choi2014,Ballance2016,Monz2016,wright2019benchmarking}.

One major problem of the linear configuration is the scalability. It is estimated that the current gate schemes can be generalized to about one hundred ions \cite{wineland1998experimental,PhysRevLett.77.3240,clark2001proceedings}, while for larger number of qubits we need more complicated schemes like ion shuttling \cite{wineland1998experimental,Kielpinski2002} or photonic quantum networks \cite{10.5555/2011617.2011618,RevModPhys.82.1209,PhysRevA.89.022317}. Although noticeable progress has been achieved in these directions in recent years, these approaches require additional exquisite control of the ion system and their speed is limited by the slow quantum wiring process.

Therefore, it is appealing to consider a scheme using the current control techniques and move the complexity into the design of the control sequence. One such possibility is to increase qubit numbers by exploring higher dimensional ion systems in 2D or 3D, where large Wigner crystals of ions have been observed in experiments for hundreds to thousands of ions \cite{PhysRevLett.96.103001,doi:10.1063/1.4705153}. Some pioneering works have been performed in Refs.~\cite{PhysRevA.90.022332,Wang2015,Bermudez_2017,PhysRevA.94.032320}. In particular, it has been shown that the micromotion of the ions, which is inevitable for large 2D or 3D ion crystals in a Paul trap and leads to deviation from the gate scheme in 1D, is not a source of decoherence. Actually, micromotion has been exploited earlier in experiments for selective single-ion addressing \cite{PhysRevLett.81.3631}. The idea is then to describe the micromotion of an ion crystal theoretically and experimentally \cite{Landa2012,Landa2012_2,PhysRevLett.109.263003,PhysRevA.92.032108}, and to find a suitable laser sequence to realize high-fidelity gates under micromotion \cite{PhysRevA.90.022332,PhysRevLett.113.053001,Wang2015,Bermudez_2017}. In previous works, approximations are made to the lowest orders of micromotion; while for high gate fidelity, it will be necessary to consider higher order corrections. In this work, we will develop an efficient algorithm to design entangling gates in general ion crystals up to arbitrary orders of micromotion. We will assume a general 3D crystal, with our results directly applicable to 2D as a special case of the trapping potential.

The paper is organized as follows. First we briefly review a commonly used gate scheme in a 1D ion chain in Sec.~\ref{sec:gate1d} and describe the effects of micromotion when generalizing it to 3D. In Sec.~\ref{sec:equilibrium} we describe a numerical method to solve the equilibrium positions and micromotion of the ions, from which we further solve all the collective normal modes in Sec.~\ref{sec:mode}. In Sec.~\ref{sec:gate} we provide efficient numerical algorithms to include the micromotion into the gate design, whose theoretical errors can be systematically controlled to arbitrary orders. A numerical example is presented in Sec.~\ref{sec:numerical}. Finally we conclude in Sec.~\ref{sec:discussion}. Appendix~\ref{app:algorithm} summarizes the numerical algorithms and Appendix~\ref{app:robustness} gives an example of incorporating the robustness condition against certain parameter drifts into our algorithms.
\section{Gate Scheme in 1D Ion Chain}
\label{sec:gate1d}
In this section, we briefly review how two-ion entangling gates can be realized in a linear chain of $N$ ions. It will later be generalized to a 3D ion crystal in the following sections. Here we will follow the notation of Ref.~\cite{PhysRevA.97.062325}.

The coupling between two ions is generated by off-resonantly driving them near some motional sidebands simultaneously. Assuming two driving fields on the ion $j$ with the same Rabi frequency $\Omega_j$, opposite detuning $\pm \mu$ to the atomic transition frequency and opposite wave vectors $\pm \Delta k$ along a transverse direction $x$, which is known as the phase-insensitive configuration \cite{Lee2005}, the qubit-phonon interaction Hamiltonian takes the form
\begin{equation}
\hat{H} = \sum_j \hbar \Omega_j \cos \left[\mu t + \varphi_j - \Delta k \cdot \hat{x}_j(t)\right] \hat{\sigma}_j^x,
\label{eq:H_phase_insensitive}
\end{equation}
where the index $j$ runs over the two ions being driven, $\varphi_j$ a motional phase depending on the initial phases of the driving fields and the equilibrium positions of the ions, $\hat{\sigma}_j^x$ the corresponding Pauli $X$ operators, and $\hat{x}_j(t)$ the time-evolved position operators in the transverse direction. We can decompose the transverse motion into normal modes
\begin{equation}
\hat{x}_j(t) = \sum_k b_j^k \sqrt{\frac{\hbar}{2 m \omega_k}} \left(\hat{a}_k e^{-i\omega_k t} + \hat{a}_k^\dag e^{i \omega_k t}\right), \label{eq:transverse_mode}
\end{equation}
where $k$ labels a normal mode with a creation (annihilation) operator $\hat{a}_k^\dag$ ($\hat{a}_k$), a mode frequency $\omega_k$, and a normalized mode vector $b_j^k$ for $j=1,\,2,\,\cdots,\,N$.

With the Lamb-Dicke parameter $\eta_k \equiv \Delta k \sqrt{\hbar/2m\omega_k} \ll 1$ for typical experiments, and assuming weak excitation of the phonon modes during the gate (which is known as the Lamb-Dicke regime), we only need to consider the lowest order expansions in $\eta_k$
\begin{equation}
\hat{H} = \sum_j \sum_k \chi_j(t) \eta_k b_j^k \left(\hat{a}_k e^{-i\omega_k t} + \hat{a}_k^\dag e^{i\omega_k t}\right) \hat{\sigma}_j^x,
\end{equation}
where
$\chi_j(t) \equiv \hbar \Omega_j \sin\left(\mu t + \varphi_j\right)$.
The zeroth order term is dropped as a single-qubit rotation which can be compensated after the entangling gate. The error of neglecting the higher order terms is shown to be $O(\eta_k^4)$ in the gate fidelity \cite{PhysRevA.97.062325}.

Time evolution under this Hamiltonian with duration $\tau$ is given by the unitary operator
\begin{equation}
\hat{U}(\tau) = \exp\left[i \sum_j \hat{\phi}_j(\tau) \hat{\sigma}_j^x + i \sum_{i<j}\Theta_{ij}(\tau)\hat{\sigma}_i^x \sigma_j^x\right],\label{eq:U_expansion}
\end{equation}
where
\begin{equation}
\hat{\phi}_j(\tau) = -i \sum_k \left[\alpha_j^k(\tau)\hat{a}_k^\dag - \alpha_j^{k*} (\tau) \hat{a}_k\right],
\end{equation}
\begin{equation}
\label{eq:alpha}
\alpha_j^k(\tau) = -\frac{i}{\hbar} \eta_k b_j^k \int_0^\tau \chi_j(t) e^{i\omega_k t} dt,
\end{equation}
describe the spin-phonon coupling after the gate, and
\begin{align}
\label{eq:phi}
&\Theta_{ij}(\tau)=\frac{1}{\hbar^2}\sum_k \eta_k^2 b_i^k b_j^k \int_0^\tau dt_1 \int_0^{t_1}dt_2 \nonumber\\
&\qquad\times\left[\chi_i(t_1) \chi_j(t_2) + \chi_j(t_1) \chi_i(t_2)\right]\sin \left[\omega_k (t_1-t_2)\right]
\end{align}
is the coupling between the two spins $i$ and $j$.

For an ideal entangling gate, we want the qubit state to be decoupled from the phonon modes, that is, $\alpha_j^k=0$; we also want $\Theta_{ij}=\pm\pi/4$ for the maximal entanglement such that the ideal gate $\exp(\pm i\pi\sigma_i^x \sigma_j^x/4)$ is equivalent to the CNOT gate up to single-qubit rotations. To characterize the deviation from the ideal gate, we compute the average gate fidelity over different initial qubit states and a thermal distribution of the phonon states. For small deviation from the ideal case, we have \cite{PhysRevA.97.062325}
\begin{equation}
\delta F = \frac{4}{5} \left[\left(\Theta_{ij}\mp \frac{\pi}{4}\right)^2 + \sum_{jk}|\alpha_j^k|^2(2 \bar{n}_k+1)\right], \label{eq:fidelity}
\end{equation}
where $\bar{n}_k$ is the average phonon number in the $k$-th mode.

Up to this point the formulation is general and we can apply amplitude, frequency or phase modulations \cite{Zhu2006,PhysRevLett.114.120502,PhysRevA.98.032318} of the driving field to optimize the gate fidelity. For concreteness, below we will focus on a specific method of amplitude modulation to optimize the gate performance.

Let us set the driving fields on the two ions to be the same and divide the gate time $\tau$ into $n_{\textrm{seg}}$ equal segments. In each segment we set the Rabi frequency on the two ions to be a constant. Define a real column vector $\boldsymbol{\Omega} = (\Omega_1, \Omega_2, \cdots, \Omega_{n_{\textrm{seg}}})^T$ corresponding to the Rabi frequency of each segment, and we get
\begin{align}
\alpha_j^k(\tau) = \boldsymbol{A}_j^k \boldsymbol{\Omega}, \quad
\Theta_{ij} = \boldsymbol{\Omega}^T \boldsymbol{\gamma}' \boldsymbol{\Omega},
\end{align}
where $\boldsymbol{A}_j^k$ is a row vector whose $n$-th component is
\begin{equation}
\label{eq:A_mat}
A_j^k (n) = -i \eta_k b_j^k \int_{(n-1)\tau/n_{\textrm{seg}}}^{n\tau/n_{\textrm{seg}}} \sin \mu t \cdot e^{i \omega_k t} dt,
\end{equation}
and $\boldsymbol{\gamma}'$ is an $n_{\textrm{seg}}$ by $n_{\textrm{seg}}$ matrix with the $(p,q)$ component given by
\begin{widetext}
\begin{equation}
\label{eq:gamma_mat}
\gamma'(p,q) = \left\{
\begin{array}{ll}
\!\!\displaystyle 2\sum_k \eta_k^2 b_i^k b_j^k \int_{(p-1)\tau/n_{\textrm{seg}}}^{p\tau/n_{\textrm{seg}}} \!\!\!\!\!\!\!\!dt_1 \int_{(q-1)\tau/n_{\textrm{seg}}}^{q\tau/n_{\textrm{seg}}} \!\!\!\!\!\!\!\!\!dt_2 \sin \mu t_1 \sin \mu t_2 \sin [\omega_k (t_1 - t_2)] & (p>q)\\
\!\!\displaystyle  2\sum_k \eta_k^2 b_i^k b_j^k \int_{(p-1)\tau/n_{\textrm{seg}}}^{p\tau/n_{\textrm{seg}}} \!\!\!\!\!\!\!\!dt_1 \int_{(p-1)\tau/n_{\textrm{seg}}}^{t_1} \!\!\!\!\!\!\!\!\!dt_2 \sin \mu t_1 \sin \mu t_2 \sin [\omega_k (t_1 - t_2)] & (p=q)\\
\!\!0 & (p<q)
\end{array}\right..
\end{equation}
\end{widetext}

By suitably scaling $\boldsymbol{\Omega}$, we can always set $\Theta_{ij} = \pm \pi/4$. Then the gate infidelity can be approximated as
\begin{equation}
\delta F = \frac{4}{5}\boldsymbol{\Omega}^T \boldsymbol{M} \boldsymbol{\Omega},\label{eq:fidelity_approx}
\end{equation}
where $\boldsymbol{M}\equiv \sum_{jk} \boldsymbol{A}_j^{k\dag} \boldsymbol{A}_j^{k}(2\bar{n}_k+1)$.
By definition, $\boldsymbol{M}$ is a Hermitian matrix, but actually we can express it in a real symmetric form since $\boldsymbol{\Omega}^T \boldsymbol{M} \boldsymbol{\Omega} = \boldsymbol{\Omega}^T \mathrm{Re}[\boldsymbol{M}] \boldsymbol{\Omega}$. Similarly we can define a symmetric matrix $\boldsymbol{\gamma} \equiv (\boldsymbol{\gamma}' + \boldsymbol{\gamma}'^T)/2$ such that $\Theta_{ij} = \boldsymbol{\Omega}^T \boldsymbol{\gamma}' \boldsymbol{\Omega} = \boldsymbol{\Omega}^T \boldsymbol{\gamma} \boldsymbol{\Omega}$.

To optimize the gate fidelity, we minimize $\boldsymbol{\Omega}^T \boldsymbol{M} \boldsymbol{\Omega}$ under the constraint $\boldsymbol{\Omega}^T \boldsymbol{\gamma} \boldsymbol{\Omega} = \pm \pi/4$. We can introduce a Lagrange multiplier and consider the optimization of $f(\boldsymbol{\Omega},\lambda)=\boldsymbol{\Omega}^T \boldsymbol{M} \boldsymbol{\Omega}-\lambda(\boldsymbol{\Omega}^T \boldsymbol{\gamma} \boldsymbol{\Omega} \mp \pi/4)$ as
\begin{equation}
\boldsymbol{M}\boldsymbol{\Omega} - \lambda \boldsymbol{\gamma} \boldsymbol{\Omega} = 0,\quad
\boldsymbol{\Omega}^T \boldsymbol{\gamma} \boldsymbol{\Omega} = \pm \pi/4.\label{eq:optimization}
\end{equation}
This is a generalized eigenvalue problem. We can solve its eigenvalue with the smallest absolute value and the corresponding eigenvector, which, after suitable normalization, gives us the optimal $\boldsymbol{\Omega}$.

Let us now briefly discuss how this gate scheme can be generalized to a general 3D ion crystal, which will be expanded in more details in the following sections.
One straightforward generalization is that we shall replace the transverse normal modes in Eq.~(\ref{eq:transverse_mode}) by all the normal modes in the 3D crystal, since the motions in different directions are coupled together.
However, there are more differences between the 1D and the 3D cases.
For the linear configuration, the trap can be designed such that the ions align on the null of the radiofrequency (RF) field without micromotion. The ions thus stay at their equilibrium positions and the small perturbation can be well approximated by the normal modes in a harmonic pseudopotential. On the other hand, in a 3D ion crystal it is generally not possible to suppress the micromotion of all the ions. Then the equilibrium solution is not a static configuration of the ions, but a finite oscillation at the RF frequency for all the ions. Such an oscillation can be absorbed into $\varphi_j$ in Eq.~(\ref{eq:H_phase_insensitive}) as a time-dependent motional phase for each ion. Moreover, when the amplitudes of these equilibrium trajectories are comparable to the width of the laser beams, the Rabi frequency $\Omega_j$ will also be time-dependent. Finally, the normal mode expansions in Eq.~(\ref{eq:transverse_mode}) will generally also include the micromotion at the RF frequency, which need to be considered in the evaluation of $\alpha_j^k$ [Eq.~(\ref{eq:alpha})] and $\Theta_{ij}$ [Eq.~(\ref{eq:phi})].
\section{Equilibrium Trajectories and Normal Modes under Micromotion}
\label{sec:equilibrium_mode}
As is mentioned above, to design the entangling gate, first we need to solve the equilibrium trajectories of the ions and the collective normal modes for small deviation away from them.
Important works on this topic have been performed in Refs.~\cite{Landa2012,Landa2012_2} which are most suitable for a few ions to the lowest orders of the micromotion. In this section, we briefly review these results and generalize them to an efficient numerical algorithm that can work for hundreds of ions and to the arbitrary orders of micromotion.

For convenience, we define the length unit $L_0=(e^2/4\pi\epsilon_0 m \omega_{\mathrm{rf}}^2)^{1/3}$ and the time unit $T_0=2/\omega_{\mathrm{rf}}$. Then we can consider the dimensionless equation of motion (EOM) of $N$ ions
\begin{align}
\label{eq:EOM2}
&\ddot{R}_{i\sigma} + \sum_\rho(A_{\sigma\rho} - 2 Q_{\sigma\rho} \cos 2t) R_{i\rho} \nonumber\\
&\qquad\qquad\qquad- 4\sum_{j\ne i} \frac{R_{i\sigma} - R_{j\sigma}}{\left[\sum_\rho(R_{i\rho}-R_{j\rho})^2\right]^{\frac{3}{2}}}=0,
\end{align}
where $i,j=1,\,2,\,\cdots,\,N$ corresponds to each ion and $\sigma,\rho=x,\,y,\,z$ for the three spatial directions. The first two terms correspond to the well-known Mathieu equation for a single ion in an RF trap, while the third term describes the Coulomb interaction between different ions with the motions in different directions coupled together. In the previous works (e.g. Refs.~\cite{RevModPhys.75.281,Landa2012}), it is usually assumed for convenience that the principal axes of the DC and the RF fields coincide, so that $A$ and $Q$ matrices are diagonal in the same frame. Since this assumption may not hold for some trap design, here we choose to work with the more general case.

\subsection{Periodic Equilibrium Solutions}
\label{sec:equilibrium}
The potential in Eq.~(\ref{eq:EOM2}) has a period of $\pi$ in the dimensionless form, it is thus reasonable to expect a solution with $\pi$ period as well \cite{Landa2012}, that is, a micromotion at the RF frequency.
To find such a stable periodic solution, we can start from some random initial configuration, time-evolve the system under a weak damping term [a $-\gamma \dot{R}_{i\sigma}$ term on the right-hand-side of Eq.~(\ref{eq:EOM2})], and gradually turn down the damping until the system reaches a stable solution.
Similar to the case of a static potential, there can be multiple stable solutions to Eq.~(\ref{eq:EOM2}). Note that the process of gradually reducing the damping term closely mimic the cooling stage of the ions in a real trap. Thus we expect the solution found in this way to reflect the one obtained in the experiment with high probability.

This method works well for a small number of ions, but difficulty appears as the ion number increases: to approach the desired solution, we need to reduce the damping term $\gamma$, which in turn slows down the convergence and requires smaller step sizes in the time-evolution to suppress the numerical errors.
This motivates us to consider the solution to Eq.~(\ref{eq:EOM2}) directly without the damping term.

Following Ref.~\cite{Landa2012}, we expand the equilibrium solution with the period of $\pi$ into a Fourier series
\begin{equation}
R_{j\sigma}^\pi(t)=\sum_{n=-\infty}^{+\infty} B_{2n,j\sigma}e^{i2nt}. \label{eq:R_fourier}
\end{equation}
Note that due to the time-reversal symmetry of Eq.~(\ref{eq:EOM2}), we have $B_{2n,j\sigma} = B_{-2n,j\sigma} = B_{2n,j\sigma}^*$. Therefore all the expansion coefficients are real.

Since $R_{i\sigma}^\pi(t)$ has a period of $\pi$, so does any function of the coordinates. Therefore the Coulomb interaction term in Eq.~(\ref{eq:EOM2}) can also be expanded into a Fourier series
\begin{align}
D_{i\sigma}^\pi(t) \equiv& 4 \sum_{j\ne i} \frac{R_{i\sigma}^{\pi}(t)-R_{j\sigma}^\pi(t)}{\left\{\sum_{\rho} \left[R_{i\rho}^{\pi}(t)-R_{j\rho}^\pi(t)\right]^2\right\}^{3/2}} \nonumber\\
=& \sum_{n=-\infty}^{+\infty} D_{2n,i\sigma} e^{i2nt}. \label{eq:D_series}
\end{align}
On the other hand, from the definition, we also have \cite{Landa2012}
\begin{equation}
D_{i\sigma}^\pi(t) = 4\sum_j G_{ij}^\pi(t) R_{j\sigma}^\pi(t),
\end{equation}
where
\begin{align}
G_{ij}^\pi(t) \equiv& \delta_{ij} \sum_{k\ne i}\frac{1}{\left\{\sum_{\rho} \left[R_{i\rho}^{\pi}(t)-R_{k\rho}^\pi(t)\right]^2\right\}^{3/2}} \nonumber\\
&- (1-\delta_{ij})\frac{1}{\left\{\sum_{\rho} \left[R_{i\rho}^{\pi}(t)-R_{j\rho}^\pi(t)\right]^2\right\}^{3/2}} \nonumber\\
=& \sum_{n=-\infty}^{+\infty} G_{2n,ij} e^{i2nt}. \label{eq:G_series}
\end{align}
These expansions are used in Ref.~\cite{Landa2012} to analyze the properties of the micromotion assuming small Mathieu parameters (elements of $\boldsymbol{A}$ and $\boldsymbol{Q}$ matrices). Here we generalize these equations to find the numerical solutions.

Plugging these expansions into Eq.~(\ref{eq:EOM2}) with a suitable division of the Coulomb interaction term into the above two alternative forms, we get a recurrence relation
\begin{align}
\label{eq:iteration}
&\sum_\rho \left[ (A_{\sigma\rho} - 4n^2\delta_{\sigma\rho}) B_{2n,i\rho} - Q_{\sigma\rho} (B_{2n-2,i\rho} +  B_{2n+2,i\rho}) \right] \nonumber\\
&\quad+ 4\alpha \sum_{m,j} G_{2n-2m,ij} B_{2m,j\sigma}= (1+\alpha)D_{2n,i\sigma},
\end{align}
where the index $m$ runs over all the orders of Fourier series, while the $\alpha$-dependent terms on the two sides cancel each other for a periodic solution.

For a desired accuracy of the solution, suppose we can truncate at some finite order of the Fourier series. Then Eq.~(\ref{eq:iteration}) is a system of linear equations for $\{B_{2n,i\sigma}\}$ and can be solved iteratively: we can use the old solution of $\{B_{2n,i\sigma}\}$ to calculate $\{D_{2n,i\sigma}\}$ and $\{G_{2n,i\sigma}\}$ and then find the new solution of $\{B_{2n,i\sigma}\}$. The simplest case is to set $\alpha=0$. Unfortunately, numerically we find that in this case the solution is unstable: if we start from an approximate solution, it will deviate further and further away after iterations.
The reason is that for the leading order term $B_{0,i\sigma}$, we have $n=0$ and thus
its coefficient is a small parameter.
When computing the new solution from the old one, we take the inverse of these small parameters and therefore the error gets enlarged.

The above analysis suggests that we need large $\alpha$ for the iterative method to converge. Numerically we find that $\alpha\ge 1$ leads to convergence around the periodic solution, if we start from the approximate solution we found with a weak damping term. Note that if in $B_{2n,i\sigma}$ we truncate at $n=\pm M$, then in $G_{2n,ij}$ we need to truncate at $n=\pm 2M$.

\subsection{Normal Modes}
\label{sec:mode}
Having solved the equilibrium trajectories $\{R_{i\sigma}^\pi(t)\}$ of the ions, now we consider small perturbation around them and decompose them into collective normal modes.
Let us first derive the equation of motion for a small deviation. Following Ref.~\cite{Landa2012}, we define a matrix function $\boldsymbol{K}(t)$ whose elements are
\begin{align}
\label{eq:def_K}
&K_{i\sigma,j\tau}(t) \nonumber\\
=& \left\{
\begin{array}{ll}
-3\frac{(R_{i\sigma}^\pi - R_{j\sigma}^\pi)(R_{i\tau}^\pi - R_{j\tau}^\pi)}{\left[\sum_\rho(R_{i\rho}^\pi - R_{j\rho}^\pi)^2\right]^{5/2}} & (i\ne j,\sigma\ne \tau)\\
\frac{\left[\sum_\rho(R_{i\rho}^\pi - R_{j\rho}^\pi)^2\right] - 3(R_{i\sigma}^\pi - R_{j\sigma}^\pi)^2}{\left[\sum_\rho(R_{i\rho}^\pi - R_{j\rho}^\pi)^2\right]^{5/2}} & (i\ne j,\sigma= \tau)\\
3\sum\limits_{k\ne i} \frac{(R_{i\sigma}^\pi - R_{k\sigma}^\pi)(R_{i\tau}^\pi - R_{k\tau}^\pi)}{\left[\sum_\rho(R_{i\rho}^\pi - R_{k\rho}^\pi)^2\right]^{5/2}} & (i= j,\sigma\ne \tau)\\
-\sum\limits_{k\ne i} \frac{\left[\sum_\rho(R_{i\rho}^\pi - R_{k\rho}^\pi)^2\right]-3(R_{i\sigma}^\pi - R_{k\sigma}^\pi)^2}{\left[\sum_\rho(R_{i\rho}^\pi - R_{k\rho}^\pi)^2\right]^{5/2}} & (i= j,\sigma= \tau)\\
\end{array}
\right.
\end{align}
as the time-dependent Hessian matrix of the Coulomb interaction.
In the above equation, the $t$ dependence of $R_{i\sigma}^\pi$ is omitted for simplicity. Now we have the linear EOM for small perturbation $r_{i\sigma}$ around the periodic crystal solution:
\begin{equation}
\label{eq:linear}
\ddot{r}_{i\sigma} + \sum_{\rho}(A_{\sigma\rho} - 2 Q_{\sigma\rho} \cos 2t)r_{i\rho} + 4 \sum_{j\rho} K_{i\sigma,j\rho}(t) r_{j\rho} = 0.
\end{equation}

Because $\boldsymbol{K}(t)$ also has a period of $\pi$, we can expand it as
\begin{equation}
\label{eq:K_expansion}
\boldsymbol{K}(t) = \boldsymbol{K}_0 - 2\boldsymbol{K}_2 \cos 2t - 2\boldsymbol{K}_4 \cos 4t - \cdots
\end{equation}
If we absorb $\boldsymbol{K}_0$ and $\boldsymbol{K}_2$ into the definition of $\boldsymbol{A}$ and $\boldsymbol{Q}$ matrices and neglect the higher order terms (we will later describe how the higher order terms can be included), Eq.~(\ref{eq:linear}) can be written in the vector form (the vectorial Mathieu equation)
\begin{equation}
\label{eq:coupled_Mathieu}
\ddot{\boldsymbol{r}} + (\boldsymbol{A} - 2 \boldsymbol{Q} \cos 2t)\boldsymbol{r} = 0,
\end{equation}
where $\boldsymbol{r}$ is a $3N$ by 1 column vector.

Consider a normal mode at the dimensionless frequency $\beta$ (modulated by micromotion at the RF frequency). Following Ref.~\cite{Landa2012}, such a solution can be expanded as
\begin{equation}
\label{eq:expansion}
\boldsymbol{r} = \sum_{n=-\infty}^{+\infty} \boldsymbol{C}_{2n}\left[c e^{i(2n+\beta)t}+c^* e^{-i(2n+\beta)t}\right].
\end{equation}
Here the real $3N$ by 1 vector $\boldsymbol{C}_{2n}$ is the generalized mode vector and $c$ is a complex amplitude.
Now our task is to determine all the mode frequencies $\beta$'s. An imaginary $\beta$ means instability, while all $\beta$'s being real corresponds to oscillatory behavior.

We define $\boldsymbol{R}_{2n} \equiv \boldsymbol{A} - (2n+\beta)^2\boldsymbol{I}$, where $\boldsymbol{I}$ is the identity matrix. Plugging these expressions into Eq.~(\ref{eq:coupled_Mathieu}), we get a recurrence relation
\begin{equation}
\label{eq:recursion}
\boldsymbol{Q} \boldsymbol{C}_{2n-2} = \boldsymbol{R}_{2n} \boldsymbol{C}_{2n} - \boldsymbol{Q} \boldsymbol{C}_{2n+2}.
\end{equation}
Further applying the infinite continued matrix inversion method \cite{Landa2012_2}, the normal mode frequencies are finally determined by
\begin{equation}
\label{eq:beta}
\det\left[\boldsymbol{P}_2(\beta) - \boldsymbol{Q} \boldsymbol{P}_1(\beta) \boldsymbol{Q}\right]=0,
\end{equation}
with $\boldsymbol{C}_0$ the eigenvector of $\boldsymbol{P}_2-\boldsymbol{Q}\boldsymbol{P}_1\boldsymbol{Q}$ corresponding to the eigenvalue of zero, while
\begin{equation}
\label{eq:continued_inversion}
\boldsymbol{P}_1(\beta)\equiv \{\boldsymbol{R}_2 - \boldsymbol{Q}[\boldsymbol{R}_4 - \boldsymbol{Q}(\boldsymbol{R}_6 - \cdots)^{-1}\boldsymbol{Q}]^{-1}\boldsymbol{Q}\}^{-1}
\end{equation}
and
\begin{equation}
\label{eq:continued_inversion_2}
\boldsymbol{P}_2(\beta) \equiv \boldsymbol{R}_0 - \boldsymbol{Q}[\boldsymbol{R}_{-2} - \boldsymbol{Q}(\boldsymbol{R}_{-4} - \cdots)^{-1}\boldsymbol{Q}]^{-1}\boldsymbol{Q}
\end{equation}
are two matrices dependent on $\beta$ and can be evaluated by truncating at some large value of $|n|$.

Note that $\det(\boldsymbol{P}_2-\boldsymbol{Q}\boldsymbol{P}_1\boldsymbol{Q})$ is just a numerical function of $\beta$. Therefore, to solve all the normal modes, in principle we only need to find all the roots of this function numerically, as done in Refs.~\cite{Landa2012,Landa2012_2}. However, for large ion number $N$, the separation between these $3N$ roots is small and it is difficult to solve all of them without any \emph{a priori} knowledge about their distribution. Moreover, some of the $\beta$'s may be imaginary for an instable crystal solution, which makes the numerical search even harder.
Also note that when writing down Eq.~(\ref{eq:coupled_Mathieu}), we have neglected higher order terms in Eq.~(\ref{eq:K_expansion}). The effect of the $K_4$ term is considered in Ref.~\cite{Landa2012_2}, but the method is difficult to generalize to higher order terms. Now we describe a new method that can be efficiently applied to a large number of ions and can include arbitrarily high order terms.

In general, we can rewrite Eq.~(\ref{eq:linear}) as
\begin{equation}
\label{eq:coupled_Mathieu_complete}
\ddot{\boldsymbol{r}} + (\boldsymbol{A} - 2 \boldsymbol{Q} \cos 2t - 2 \boldsymbol{Q}_4 \cos 4t -\cdots)\boldsymbol{r} = 0
\end{equation}
with $\boldsymbol{Q}_{2n}=4\boldsymbol{K}_{2n}$ ($n\ge 2$) from the Fourier expansion of $\boldsymbol{K}(t)$ [Eq.~(\ref{eq:K_expansion})]. Again we have absorbed $\boldsymbol{K}_0$ and $\boldsymbol{K}_2$ into the definition of $\boldsymbol{A}$ and $\boldsymbol{Q}$.

Using the expansion of Eq.~(\ref{eq:expansion}), we get a general recurrence relation
\begin{align}
\boldsymbol{R}_{2n}\boldsymbol{C}_{2n} =& \boldsymbol{Q}(\boldsymbol{C}_{2n-2}+\boldsymbol{C}_{2n+2})
+ \boldsymbol{Q}_4(\boldsymbol{C}_{2n-4}+\boldsymbol{C}_{2n+4}) \nonumber\\
&\qquad+ \boldsymbol{Q}_6(\boldsymbol{C}_{2n-6}+\boldsymbol{C}_{2n+6}) +\cdots
\end{align}
where again $\boldsymbol{R}_{2n}\equiv \boldsymbol{A}-(2n+\beta)^2\boldsymbol{I}$.

If we truncate these equations at the order of $\pm n$, we can assemble them into a matrix form
\begin{widetext}
\begin{equation}
\left(
\begin{array}{ccccccc}
\boldsymbol{R}_{-2n} & -\boldsymbol{Q} & -\boldsymbol{Q}_4 & \cdots & 0 & 0 & 0 \\
-\boldsymbol{Q} & \boldsymbol{R}_{-2n+2} & -\boldsymbol{Q} & \cdots & 0 & 0 & 0 \\
-\boldsymbol{Q}_4 & -\boldsymbol{Q} & \boldsymbol{R}_{-2n+4} & \cdots & 0 & 0 & 0 \\
\vdots & \vdots & \vdots & \ddots & \vdots & \vdots & \vdots \\
0 & 0 & 0 & \cdots & \boldsymbol{R}_{2n-4} & -\boldsymbol{Q} & -\boldsymbol{Q}_4\\
0 & 0 & 0 & \cdots & -\boldsymbol{Q} & \boldsymbol{R}_{2n-2} & -\boldsymbol{Q}\\
0 & 0 & 0 & \cdots & -\boldsymbol{Q}_4 & -\boldsymbol{Q} & \boldsymbol{R}_{2n}
\end{array}
\right) \left(
\begin{array}{c}
\boldsymbol{C}_{-2n}\\
\boldsymbol{C}_{-2n+2}\\
\boldsymbol{C}_{-2n+4}\\
\vdots\\
\boldsymbol{C}_{2n-4}\\
\boldsymbol{C}_{2n-2}\\
\boldsymbol{C}_{2n}\\
\end{array}
\right) = 0, \label{eq:mode_matrix_iterative}
\end{equation}
\end{widetext}
where $\boldsymbol{Q}_{2n}$'s can be truncated at some different order $m$, which is not shown
explicitly in the above equation. Note that the diagonal blocks $\{\boldsymbol{R}_{2n}\}$ of this matrix depend on $\beta$. Again a normal mode frequency $\beta$
is a root of the determinant of this matrix, and the mode vector is given by the corresponding eigenvector
with the eigenvalue of zero.

The direct search of $\beta$ suffers the same numerical difficulty as mentioned before. Instead, we observe that if we have an approximate mode frequency $\beta$, the above matrix should have an eigenvalue close to zero. Therefore, we start from an approximate solution and solve the eigenvalue $\Delta$ closest to zero. Note that the diagonal blocks of this matrix $\boldsymbol{R}_{2n} = \boldsymbol{A}-(2n+\beta)^2\boldsymbol{I} = \boldsymbol{A} - (4n^2 +4n\beta)\boldsymbol{I} - \beta^2\boldsymbol{I}$ have a common $-\beta^2\boldsymbol{I}$ term in the diagonal, which can be combined with the small nonzero eigenvalue $\Delta$. That is, we update $\beta$ by $\sqrt{\beta^2+\Delta}$ and then repeat the above process. The iterations will quickly converge such that $\Delta\to 0$. In the case of degeneracy of, say, two normal modes, we can similarly look for the two eigenvalues with the smallest absolute values and the corresponding eigenvectors, and use the desired one for the next iteration.

To efficiently apply the above method, it is desirable to have a complete set of approximate normal modes to start with. Therefore here we describe a simple approximate solution based on Eq.~(\ref{eq:beta}). Consider small elements of the $\boldsymbol{A}$ and $\boldsymbol{Q}$ matrices [after absorbing $\boldsymbol{K}_0$ and $\boldsymbol{K}_2$ as in Eq.~(\ref{eq:coupled_Mathieu})], and we expect the mode frequencies $\beta$ also to be small.
Now we consider the following expansion for $n\ne 0$:
\begin{align}
\boldsymbol{R}_{2n}^{-1} =& \left[\boldsymbol{A}-(2n+\beta)^2\right]^{-1} = \left(\boldsymbol{A}-4n^2-4n\beta-\beta^2\right)^{-1}\nonumber\\
\approx&-\frac{1}{4n^2}\left(1 - \frac{\beta}{n}-\frac{\beta^2-\boldsymbol{A}}{4n^2}+\frac{\beta^2}{n^2}\right),
\end{align}
where we have omitted the identity matrix $\boldsymbol{I}$ for simplicity.
We also have
\begin{align}
\boldsymbol{P}_1\approx\left[\boldsymbol{R}_2 - \boldsymbol{Q} \boldsymbol{R}_4^{-1} \boldsymbol{Q}\right]^{-1} \approx \boldsymbol{R}_2^{-1} + \boldsymbol{R}_2^{-1} \boldsymbol{Q} \boldsymbol{R}_4^{-1} \boldsymbol{Q} \boldsymbol{R}_2^{-1},
\end{align}
and
\begin{align}
\boldsymbol{P}_2\approx& \boldsymbol{R}_0 - \boldsymbol{Q} \left[\boldsymbol{R}_{-2} - \boldsymbol{Q} \boldsymbol{R}_{-4}^{-1} \boldsymbol{Q}\right]^{-1} \boldsymbol{Q} \nonumber\\
\approx& \boldsymbol{R}_0 - \boldsymbol{Q} \boldsymbol{R}_{-2}^{-1} \boldsymbol{Q} - \boldsymbol{Q} \boldsymbol{R}_{-2}^{-1} \boldsymbol{Q} \boldsymbol{R}_{-4}^{-1} \boldsymbol{Q} \boldsymbol{R}_{-2}^{-1} \boldsymbol{Q}.
\end{align}
Therefore
\begin{align}
\boldsymbol{T}\equiv &\boldsymbol{P}_2 - \boldsymbol{Q} \boldsymbol{P}_1 \boldsymbol{Q} \nonumber\\
\approx& \boldsymbol{A} + \frac{1}{2}\boldsymbol{Q}^2 + \frac{1}{8}\boldsymbol{Q} \boldsymbol{A} \boldsymbol{Q} + \frac{1}{128}\boldsymbol{Q}^4 - \beta^2\left(1-\frac{3}{8}\boldsymbol{Q}^2\right).
\end{align}

Because $\boldsymbol{C}_0$ satisfies $\boldsymbol{T}\boldsymbol{C}_0=0$, all the normal modes are now given by a generalized eigenvalue problem
\begin{align}
&\left(\boldsymbol{A} + \frac{1}{2}\boldsymbol{Q}^2 + \frac{1}{8}\boldsymbol{Q} \boldsymbol{A} \boldsymbol{Q} + \frac{1}{128}\boldsymbol{Q}^4\right)\boldsymbol{C}_0
=\nonumber\\
&\qquad\qquad\qquad\qquad\qquad\qquad \beta^2 \left(1-\frac{3}{8}\boldsymbol{Q}^2\right) \boldsymbol{C}_0. \label{eq:generalized_eigen}
\end{align}
To describe the normal modes to the same order of approximation we include oscillations at the multiples of the RF frequency, which are given by
\begin{align}
\boldsymbol{C}_{\pm 2} =& \left(\boldsymbol{R}_{\pm 2} - \boldsymbol{Q} \boldsymbol{R}_{\pm 4}^{-1} \boldsymbol{Q}\right)^{-1} \boldsymbol{Q} \boldsymbol{C}_0 \approx -\frac{1}{4}(1\mp\beta) \boldsymbol{Q}\boldsymbol{C}_0,\label{eq:C_approx_2nd}\\
\boldsymbol{C}_{\pm 4} =& \boldsymbol{R}_{\pm 4}^{-1} \boldsymbol{Q} \boldsymbol{C}_{\pm 2} \approx \frac{1}{64}\left(1\mp \frac{3}{2}\beta\right)\boldsymbol{Q}^2\boldsymbol{C}_0\label{eq:C_approx_4th}.
\end{align}

Finally, we want to mention that we have been considering quadrupole traps in the above derivations (represented by the $\boldsymbol{A}$ and $\boldsymbol{Q}$ matrices). By exploring symmetries in the trap design, octupole or even higher-order multipole traps can be achieved, which have been used to trap large ion crystals \cite{PhysRevA.80.043405}. These trapping forces oscillating at the multiples of the RF frequency can be treated in a similar way as the nonlinear Coulomb interaction in the above expressions and therefore our method is still applicable.

\subsection{Quantization of normal modes}
Having solved all the $3N$ normal mode frequencies $\beta_k$ and the mode vectors $\boldsymbol{C}_{2n}^{(k)}$, finally we want to quantize them to design the entangling gate. In this part we recover the dimensions of the variables. In analogue to the quantum harmonic oscillators, we can decompose the position and momentum operators of the $N$ ions as \cite{RevModPhys.75.281}
\begin{equation}
\hat{\boldsymbol{r}}(t) = \sum_k \sqrt{\frac{\hbar}{2m\omega_k}}\left[ \hat{a}_k \boldsymbol{u}^{(k)*}(t) + \hat{a}_k^\dag \boldsymbol{u}^{(k)}(t) \right],
\label{eq:r_expansion}
\end{equation}
and
\begin{equation}
\hat{\boldsymbol{p}}(t) = \sum_k \sqrt{\frac{\hbar m}{2\omega_k}}\left[ \hat{a}_k \dot{\boldsymbol{u}}^{(k)*}(t) + \hat{a}_k^\dag \dot{\boldsymbol{u}}^{(k)}(t) \right],
\label{eq:p_expansion}
\end{equation}
where
\begin{equation}
\boldsymbol{u}^{(k)}(t) = \sum_{n=-\infty}^{+\infty} \boldsymbol{C}_{2n}^{(k)} e^{i(n\omega_{\mathrm{rf}}+\omega_k)t}
\end{equation}
is a solution to Eq.~(\ref{eq:linear}) we get in Sec.~\ref{sec:mode} with the time dimension recovered. $\omega_k=\beta_k\omega_{\mathrm{rf}}/2$ is the frequency of the $k$-th mode and $\hat{a}_k$ and $\hat{a}_k^\dag$ are the annihilation and creation operators.
What remains to be determined is the normalization of $\boldsymbol{C}_{2n}^{(k)}$, which is fixed by the commutation relations $[\hat{r}_{i\sigma}(0),\hat{p}_{j\rho}(0)]=i\hbar\delta_{ij}\delta_{\sigma\rho}$ and $[\hat{a}_k,\hat{a}_l^\dag]=\delta_{kl}$. According to Ref.~\cite{Landa2012_2}, we need
\begin{equation}
\sum_n (2n+\beta_k) \boldsymbol{C}_{2n}^{(k)T} \sum_m \boldsymbol{C}_{2m}^{(l)} = \beta_k\delta_{kl}. \label{eq:orthonormal}
\end{equation}
Actually, the orthogonal condition is already satisfied from our solution in Sec.~\ref{sec:mode}; here we just need $\sum_n (2n+\beta_k) \boldsymbol{C}_{2n}^{(k)T} \sum_m\boldsymbol{C}_{2m}^{(k)} = \beta_k$ ($k=1,\,2,\,\cdots,\,3N$) for normalization.

\section{Gate Scheme in 3D}
\label{sec:gate}
Having solved the equilibrium trajectories and the normal modes of the ions, now we can consider the gate design. Recall that, without micromotion, the optimization of the entangling gate is achieved by minimizing the residual entanglement to the phonon modes [Eq.~(\ref{eq:alpha})]
\begin{equation}
\label{eq:alpha3}
\alpha_j^k(\tau) = -\frac{i}{\hbar} \eta_k \int_0^\tau b_j^k \chi_j(t) e^{i\omega_k t} dt,
\end{equation}
while maintaining the desired two-qubit phase $\Theta_{ij}=\pm \pi/4$ [Eq.~(\ref{eq:phi})]
\begin{align}
\label{eq:phi_complex3}
\Theta_{ij}=&\frac{1}{\hbar^2}\sum_k \eta_k^2 \cdot \mathrm{Im} \Bigg\{\int_0^\tau \!\!\!dt_1 \int_0^{t_1}\!\!\!dt_2 e^{i\omega_k(t_1-t_2)} \nonumber\\
&\times \Big[b_i^k b_j^{k*}\chi_i(t_1) \chi_j(t_2) + b_j^k b_i^{k*} \chi_j(t_1) \chi_i(t_2)\Big] \Bigg\},
\end{align}
where we have considered the possibility of a complex mode vector $b_j^k$ for reasons that will become clear later, and
\begin{equation}
\label{eq:chi3}
\chi_j(t) \equiv \hbar \Omega_j(t) \sin\left(\mu t + \varphi_j\right)
\end{equation}
describes the driving field felt by the ion $j$.

In Sec.~\ref{sec:gate1d} we consider piecewise-constant $\Omega_j(t)$. The advantage is that the time integrations above can be performed analytically on each segment. Then the optimization becomes a generalized eigenvalue problem [Eq.~(\ref{eq:optimization})], which can be solved efficiently.
Note that, strictly speaking, an analytical expression for the time integral is not necessary in the above process. Given a gate time $\tau$ and the number of segments $n_{\mathrm{seg}}$, we can also evaluate the integral numerically.
However, for the highly oscillatory functions we are considering, such a numerical integration requires very high accuracy and is usually slow. In comparison, if we can derive an analytical expression for the integral, its evaluation becomes much faster, which allows efficient optimization of the gate design. This understanding is important for efficient gate design in the general case.

\subsection{Effects of micromotion}
Now we consider the effects of the micromotion on the gate design. As briefly discussed in Sec.~\ref{sec:gate1d}, it appears in the following three aspects.

\emph{Time-Dependent Motional Phase.} As we have shown in Sec.~\ref{sec:equilibrium}, the equilibrium trajectories of the ions are oscillating at the RF frequency. In Eq.~(\ref{eq:H_phase_insensitive}), it appears as a (classical) oscillating phase in the form of $\boldsymbol{\Delta k}\cdot \boldsymbol{R}_j(t)$ where $\boldsymbol{R}_j(t)=[R_{jx}^\pi(t),\,R_{jy}^\pi(t),\,R_{jz}^\pi(t)]^T$ is the equilibrium trajectory of the ion $j$. We can absorb it into the motional phase $\varphi_j$ for each ion in Eq.~(\ref{eq:chi3}) with a period of $2\pi/\omega_{\mathrm{rf}}$. Hence we have the Fourier expansion
\begin{equation}
\varphi_j(t) = \sum_{l=0}^{\infty} \varphi_j^{(l)}\cos l\omega_{\mathrm{rf}}t, \label{eq:motional_phi_expansion}
\end{equation}
where the superscript $l$ represents the $l$-th order of expansion. Note that by definition $\varphi_j(t)$ follows the same time-reversal symmetry as $R_{j\sigma}^\pi(t)$, so in the Fourier series only the cosine terms survive. Also, $\varphi_j(t)$'s depend not only on the amplitude of the micromotion, but also on its angle with the direction of $\boldsymbol{\Delta k}$.

\emph{Time-Dependent Complex Normal Modes.}
The micromotion also leads to high-frequency modulation on each normal mode, as shown in Sec.~\ref{sec:mode}. From Eq.~(\ref{eq:transverse_mode}) and Eq.~(\ref{eq:r_expansion}) we observe that $b_j^k e^{i\omega_k t}$ is now replaced by $\boldsymbol{u}^{(k)}(t)$, that is,
\begin{equation}
b_j^k \to \sum_{n=-\infty}^{+\infty} \boldsymbol{m} \cdot \boldsymbol{C}_{2n,j}^{(k)} e^{i n\omega_{\mathrm{rf}}t} \label{eq:normal_mode_time_dependent}
\end{equation}
where $\boldsymbol{m}$ is a unit vector along the direction of $\boldsymbol{\Delta k}$ and $\boldsymbol{C}_{2n,j}^{(k)}=[C_{2n,jx}^{(k)},\,C_{2n,jy}^{(k)},\,C_{2n,jz}^{(k)}]^T$.
This gives a time-dependent multiplicative factor in the integration in $\alpha_j^k$ [Eq.~(\ref{eq:alpha3})] and $\Theta_{ij}$ [Eq.~(\ref{eq:phi_complex3})], but because it is a trigonometric function, an analytical expression is still easy to obtain. Also note that our expression for $\Theta_{ij}$ in Sec.~\ref{sec:gate1d} [Eq.~(\ref{eq:phi})] is derived for a real mode vector. For the complex mode vector here, Eq.~(\ref{eq:phi_complex3}) should be used instead.

\emph{Time-Dependent Modulation of Laser Intensity.}
If the micromotion has a component perpendicular to the direction of the laser beam, it will also lead to a variation in the laser intensity felt by the ion during the gate. This become important if the amplitude of the micromotion is greater than or comparable to the width of the laser beam. Mathematically it is described by a time-dependent effective Rabi frequency $\Omega_j(t)$ in Eq.~(\ref{eq:chi3}) oscillating at the RF frequency. Given the spatial power distribution of the driving laser and the equilibrium trajectories of the ions, these terms can be computed and again can be expanded into a Fourier series. Therefore again they are multiplicative trigonometric functions and can be treated in the same way as the time-dependent normal modes. We will not consider this effect below for simplicity, which corresponds to a small micromotion amplitude compared with the width of the driving laser beam for individual addressing, or a beam that follows the equilibrium micromotion of the ions.

\subsection{Series Expansion for Evaluating Highly Oscillatory Functions}
\label{sec:micromotion_numerical}
As mentioned above, to efficiently design the gate, we want analytical expressions for Eq.~(\ref{eq:alpha3}) and Eq.~(\ref{eq:phi_complex3}) with piecewise constant $\Omega_j(t)$ on arbitrary intervals.

The last two effects of micromotion we described are not difficult to treat because for them the integrand is still the product of trigonometric functions, and therefore simple analytical expressions exist.
On the other hand, the first effect of a time-dependent motional phase, even if only expanded to the first order, will give us something like $\cos(\varphi^{(1)}\cos\omega_{\mathrm{rf}}t)$ whose analytical integral on a general time interval is not known.
Without such an analytical expression, we will have to numerically integrate highly oscillating functions, which significantly increases the computational cost.

As hinted by Ref.~\cite{Condon2009}, this problem can be solved by a series expansion of the motional phase. For each expansion term we can still derive an analytical expression, while the overall error can be systematically suppressed by considering higher order terms. At the essence of our algorithm is the following formula \cite{arfken2013mathematical}
\begin{equation}
\label{eq:Bessel}
\exp(i\varphi\cos\omega t) = J_0(\varphi) + 2\sum_{n=1}^{\infty} i^n J_n(\varphi) \cos (n\omega t).
\end{equation}
Note that for $0<\varphi\ll\sqrt{n+1}$, we have
\begin{equation}
\label{eq:Bessel_asymptotic}
J_n(\varphi) \sim \frac{1}{n!} \left(\frac{\varphi}{2}\right)^n,
\end{equation}
thus the high-order terms vanish quickly as $n$ increases, and we can expect a fast convergence.

\subsection{Residual Spin-Phonon Coupling}
\label{sec:alpha_expansion}
For the $\alpha_j^k$ terms [Eq.~(\ref{eq:alpha3})], we want to derive an analytical expression for the integral
\begin{equation}
\int \!\! dt \!\! \sum_{n=-\infty}^{+\infty}\!\! C_{2n,j\sigma}^{(k)} e^{i n\omega_{\mathrm{rf}}t} \sin \left( \mu t + \sum_{l=0}^{\infty} \varphi_j^{(l)}\cos l\omega_{\mathrm{rf}}t\right) e^{i \omega_k t}, \label{eq:50}
\end{equation}
Because $\boldsymbol{C}_{2n}^{(k)}$ decreases quickly with $|n|$ (for small Mathieu $q$ parameters, that is, elements of the $\boldsymbol{Q}$ matrix), we can truncate the summation over $n$ at small $\pm n_{\mathrm{cut}}$.

Note that we can regard $n\omega_{\mathrm{rf}}+\omega_k$ as a new variable $\omega$ in the above expression. Then once we derive an analytical expression as a function of $\omega$, we immediately get the results for all the orders of $n$ by plugging in the value of $n\omega_{\mathrm{rf}}+\omega_k$. Therefore we only need to consider
\begin{align}
&\int dt \sin \left( \mu t + \sum_{l=0}^{\infty} \varphi_j^{(l)}\cos l\omega_{\mathrm{rf}}t\right) e^{i \omega t}\nonumber\\
=& \frac{1}{2i}\int dt \left[e^{i(\mu t + \varphi_j^{(0)})}\prod_{l=1}^{\infty} e^{ i \varphi_j^{(l)}\cos l\omega_{\mathrm{rf}}t} - c.c.\right] e^{i \omega t}.
\label{eq:expand_single}
\end{align}

We can now expand $\exp(i \varphi_j^{(l)}\cos l\omega_{\mathrm{rf}}t)$ into cosine functions using Eq.~(\ref{eq:Bessel}). Suppose the $\varphi_j^{(1)},\,\varphi_j^{(2)},\,\cdots$ terms are expanded to the order of $n_1,\,n_2,\,\cdots$, respectively. Then one term of the integral becomes
\begin{equation}
\label{eq:cos_prod}
\int dt e^{i(\mu t + \varphi_j^{(0)})} e^{i \omega t} \cos (n_1\omega_{\mathrm{rf}}t) \cos ( 2 n_2\omega_{\mathrm{rf}}t) \cdots
\end{equation}
with the coefficient $\prod_l 2i^{n_l} J_{n_l}(\varphi_j^{(l)})$ (for a term with $n_l=0$ the coefficient is smaller by one half). Then we sum over all possible $\{n_1,\,n_2,\,\cdots\}$ to get the total integral.
The other half of Eq.~(\ref{eq:expand_single}) for the complex conjugate can be computed similarly. We just need to replace $\mu$ and $\varphi_j^{(0)}$ in Eq.~(\ref{eq:cos_prod}) by $-\mu$ and $-\varphi_j^{(0)}$, and to take the complex conjugate of the corresponding coefficients.

It seems that the number of terms to be evaluated is exponentially large, so that even if we have analytical expressions for Eq.~(\ref{eq:cos_prod}) and the evaluation of a single term is fast,
the overall time cost is still high. Fortunately, in the small $q$ regime typical for current ion trap quantum computing experiments, the micromotion amplitude decreases quickly with the order of expansion, so that $\varphi_j^{(l)}$ will be close to zero for large $l$.
For such terms, we have $\exp(i\varphi_{j}^{(l)}\cos l\omega_{\mathrm{rf}}t)\approx 1$, so that discarding them has negligible effects on the integrand of Eq.~(\ref{eq:expand_single}).
Therefore we can truncate at $l\le L$ for some small $L$ in Eq.~(\ref{eq:cos_prod}). Even for the remaining $\varphi_j^{(l)}$ terms, the corresponding coefficients $J_{n_l}(\varphi_j^{(l)})$ quickly vanish for large $n_l$, so the number of terms we need to evaluate is small. Note that all the truncations described above can be performed according to a chosen error tolerance, say, $10^{-8}$, and the accuracy can be systematically improved by including higher order terms.

What remains is to evaluate Eq.~(\ref{eq:cos_prod}) for up to $L$ cosine functions.
For a given $L$, an analytical expression for this integral is possible, but it can be complicated even for moderate $L$ and in particular can be difficult for coding. Therefore we further split each cosine function into two exponential functions using $\cos x=(e^{ix}+e^{-ix})/2$. In this way, Eq.~(\ref{eq:cos_prod}) finally turns into at most $2^L$ integrals of exponential functions, whose analytical expressions are simple. Since the truncation $L$ is typically small and we do not need to split for the $l$-th term if $n_l=0$, this algorithm gives a good balance between the time cost of programming and that of running.
\subsection{Two-Qubit Rotation Angle}
\label{sec:phi_expansion}
For the $\Theta_{ij}$ terms [Eq.~(\ref{eq:phi_complex3})], we want to derive an analytical expression for
\begin{align}
&\int \!\!dt_1\!\! \int\!\! dt_2 e^{i\omega_k(t_1-t_2)} \!\!\!\!\!\! \sum_{n,m=-\infty}^{+\infty}\!\!\!\! C_{2n,i\rho}^{(k)} e^{i n \omega_{\mathrm{rf}} t_1}
C_{2m,j\sigma}^{(k)} e^{-i m \omega_{\mathrm{rf}} t_2} \nonumber\\
& \qquad\qquad\qquad\times
\sin \left( \mu t_1 + \sum_{l_1=0}^{\infty} \varphi_i^{(l_1)}\cos l_1\omega_{\mathrm{rf}}t_1\right) \nonumber\\
&\qquad\qquad\qquad\times \sin \left( \mu t_2 + \sum_{l_2=0}^{\infty} \varphi_j^{(l_2)}\cos l_2\omega_{\mathrm{rf}}t_2\right) \label{eq:53}
\end{align}
for one term in Eq.~(\ref{eq:phi_complex3}), and the other term can be obtained by exchanging $i$ and $j$.

Recall that in Eq.~(\ref{eq:gamma_mat}), we need two types of integral limits
\begin{equation}
\int_{(p-1)\tau/n_{\textrm{seg}}}^{p\tau/n_{\textrm{seg}}} \!\!\!\!\!\!\!dt_1 \int_{(q-1)\tau/n_{\textrm{seg}}}^{q\tau/n_{\textrm{seg}}} \!\!\!\!\!\!\!dt_2
\quad \!\! \mathrm{and}
\int_{(p-1)\tau/n_{\textrm{seg}}}^{p\tau/n_{\textrm{seg}}} \!\!\!\!\!\!\!dt_1
\int_{(p-1)\tau/n_{\textrm{seg}}}^{t_1} \!\!\!\!\!\!\!dt_2.
\end{equation}
For the first type, the integrations over $t_1$ and $t_2$ are separable and are exactly what we have solved for the $\alpha_j^k$ terms. The remaining problem is just the second type.

Following the derivations for the $\alpha_j^k$ terms, we define new variables
$\omega_1=\omega_k+n\omega_{\mathrm{rf}}$ and $\omega_2=\omega_k+m\omega_{\mathrm{rf}}$, split the sine function into two exponential functions, and expand $\exp(\pm i \varphi_j^{(l)}\cos l\omega_{\mathrm{rf}}t)$ using Eq.~(\ref{eq:Bessel}). Now we want an analytical expression for
\begin{align}
&\int dt_1 \int dt_2 e^{\pm i(\mu t_1+\varphi_i^{(0)})} e^{\pm i(\mu t_2+\varphi_j^{(0)})} e^{i\omega_1 t_1} e^{-i\omega_2 t_2} \nonumber\\
&\qquad\times\prod_{l_1=1}^{L_1} \cos (l_1 n_{l_1}\omega_{\mathrm{rf}}t_1) \prod_{l_2=1}^{L_2} \cos (l_2 m_{l_2}\omega_{\mathrm{rf}}t_2). \label{eq:theta_integral}
\end{align}

Again we truncate at large values of $L_1$, $L_2$ and $n_{l_1}$, $m_{l_2}$ in the expansions and express the cosine functions as the sum of two exponential functions, then analytical formulae for the remaining integrals are simple.

\section{Numerical Results}
\label{sec:numerical}
\begin{figure}[!tbp]
  \centering
  \includegraphics[width=0.9\linewidth]{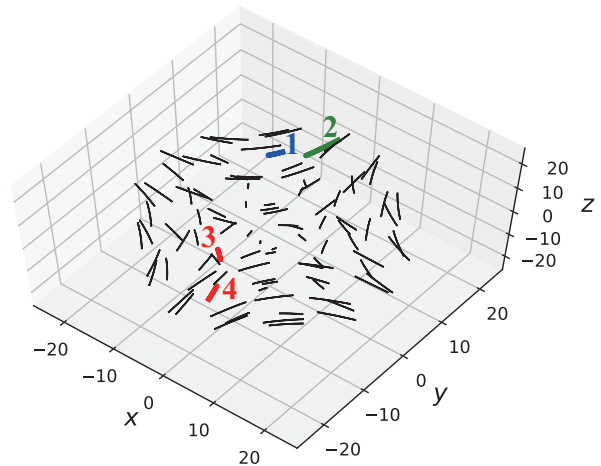}\\
  \caption{Equilibrium trajectories of 100 ions in a trap with $\boldsymbol{a}=(-0.015,\,-0.015,\,0.03)$ and $\boldsymbol{q}=(0.3,\,-0.3,\,0)$. The length dimension is $L_0=(e^2/4\pi\epsilon_0 m \omega_{\mathrm{rf}}^2)^{1/3}=0.20\,\mu$m. The two ions colored in blue (labelled by 1) and in green (labelled by 2) are used in Fig.~\ref{fig:normal_mode} to compare the solved normal modes with direct molecular dynamics simulation. The two ions colored in red (labelled by 3 and 4) are used in Fig.~\ref{fig:scan_fidelity_micro} for entangling gates.}\label{fig:equilibrium}
\end{figure}

Finally, we show a numerical example for the whole process from solving the dynamics of the ions to the gate design. Consider 100 ${}^{171}\mathrm{Yb}^+$ ions in a trap with trapping parameters $\boldsymbol{a}=(-0.015,\,-0.015,\,0.03)$ and $\boldsymbol{q}=(0.3,\,-0.3,\,0)$ (which correspond to diagonal $\boldsymbol{A}$ and $\boldsymbol{Q}$ matrices), and an RF frequency $\omega_{\mathrm{rf}}=2\pi\times 50\,$MHz. The equilibrium trajectories are plotted in Fig.~\ref{fig:equilibrium} using the numerical methods of Sec.~\ref{sec:equilibrium}. As we can see, in general the ions have large micromotion amplitudes comparable to their separations.

\begin{figure}[!tbp]
  \centering
  \includegraphics[width=0.95\linewidth]{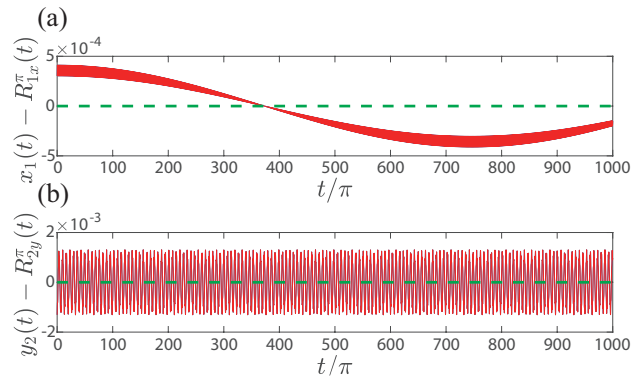}\\
  \caption{Comparison between the computed normal modes and the numerical results from molecular dynamics simulation. The 100-ion crystal we consider is shown in Fig.~\ref{fig:equilibrium}. The length unit is $L_0=0.20\,\mu$m and each dimensionless RF period $\pi$ corresponds to $0.02\,\mu$s. (a) Evolution of the $x$ coordinate of ion 1 (colored in blue in Fig.~\ref{fig:equilibrium}) over 1000 RF periods for a weak excitation of 0.01 in the lowest mode with $\beta_1=0.001340$. (b) Evolution of the $y$ coordinate of ion 2 (colored in green in Fig.~\ref{fig:equilibrium}) over 1000 RF periods for a weak excitation of 0.01 in the highest mode with $\beta_{300}=0.3032$. In both plots, the blue curve is from direct molecular dynamics simulation using a fourth order symplectic integrator (see, e.g., Ref.~\cite{thijssen_2007}), 1000 steps per RF period and \emph{double} precision, the red curve is computed from the normal mode expansions, and the green dashed line is their difference. The blue and the red curves almost coincide with each other and can hardly be distinguished in these plots, which suggests that our computed normal modes are accurate for a wide range of spectra over two orders of magnitude. The equilibrium trajectories $R_{i\sigma}^{\pi}(t)$ oscillating at an amplitude of the order $O(1)$ is already subtracted from these curves to highlight the small deviation [Eq.~(\ref{eq:linear})].}\label{fig:normal_mode}
\end{figure}

Using the methods described in Sec.~\ref{sec:mode}, we solve all the collective normal modes of the ion crystal under micromotion. Instead of presenting all the 300 modes, in Fig.~\ref{fig:normal_mode} we plot the response of a particular ion along a particular direction, when two normal modes, the lowest one at $\beta_1=0.001340$ and the highest one at $\beta_{300}=0.3032$, are selectively excited. In both cases, we see perfect agreement between the prediction of the normal mode expansion (red curves) and the direct numerical results from molecular dynamics (MD) simulation (blue curves), such that their difference (green dashed lines) stay at zero during the simulated time periods. Note that the full MD simulation results correspond to the periodic equilibrium trajectories plus the normal mode expansions, hence the good agreement in Fig.~\ref{fig:normal_mode} indicates that both our numerical methods in Sec.~\ref{sec:equilibrium} and in Sec.~\ref{sec:mode} are of high accuracy.

\begin{figure}[!tbp]
  \centering
  \includegraphics[width=0.95\linewidth]{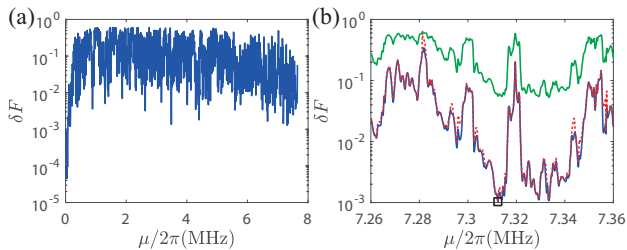}\\
  \caption{(a) Optimal gate infidelity $\delta F$ for a scan of the laser detuning $\mu$ over the whole spectrum of the normal modes of the crystal. Note that this is a coarse scan and the resolution is not enough to show all the structures of the curve. We then zoom in into a region with potential high fidelity. The calculation is performed for a truncation of $L=5$ for the equilibrium trajectories and $n_{\mathrm{cut}}=5$ for the high-frequency modulation of the normal modes. (b) A zoomed-in scan for a small range of frequencies at the high-frequency end. (We avoid using the low-frequency end, which may be sensitive to the soft modes, even though it seems to have higher theoretical fidelity.) The blue solid curve is optimized for ions' motions truncated at $L=5$ and $n_{\mathrm{cut}}=5$. The red dashed curve is what we would have gotten using the same optimized laser sequence if we truncated at $L=1$ and $n_{\mathrm{cut}}=1$, while the upper green curve is that for $L=0$ and $n_{\mathrm{cut}}=0$, that is, without considering micromotion.}\label{fig:scan_fidelity_micro}
\end{figure}

\begin{figure}[!tbp]
  \centering
  \includegraphics[width=0.8\linewidth]{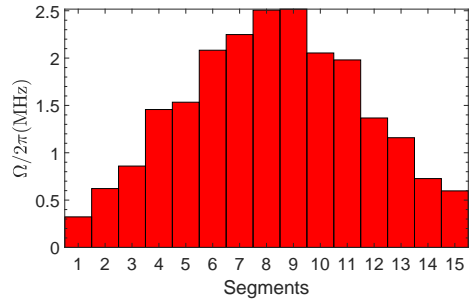}\\
  \caption{The optimized laser pulse sequence $\Omega(t)$ for $\tau=300\,\mu$s, $n_{\mathrm{seg}}=15$, $\mu=2\pi\times7.3124\,$MHz.}\label{fig:gate}
\end{figure}

Next we consider the entangling gate between two ions colored in red (labelled as 3 and 4) in Fig.~\ref{fig:equilibrium}.
Similar to Ref.~\cite{PhysRevA.97.062325}, we consider counter-propagating Raman laser beams with a wavelength around $\lambda=355\,$nm in the $x$ direction, and we assume Doppler temperature $k_B T = \hbar \Gamma/2$ where $\Gamma=2\pi\times 20\,$MHz is the spontaneous emission rate of the ${}^{171}\mathrm{Yb}^+$ ions. Note that in real experiments we may need the two Raman beams to be at an angle to selectively address individual ions at their intersection. For convenience, we set the static motional phase $\varphi_j^{(0)}$ to zero, which can be realized by suitable phase shift on the path of the laser beams, and only consider the oscillating motional phase $\varphi_j^{(l)}$ ($l\ge 1$) due to the micromotion. Suppose we use $n_{\mathrm{seg}}=15$ segments for a total gate time $\tau=300\,\mu$s. The optimal gate infidelity is shown in Fig.~\ref{fig:scan_fidelity_micro} as we scan the laser detuning $\mu$. Due to the increased computational cost under micromotion and the large range of $\mu$ to be scanned over, first we perform a coarse scan in Fig.~\ref{fig:scan_fidelity_micro}(a) and then a finer scan in the region with potential high gate fidelity in Fig.~\ref{fig:scan_fidelity_micro}(b). It seems that higher fidelity is possible in the low-frequency end, but here we purposely avoid this region, which may strongly drive the low-frequency soft modes and break down the approximations in the gate scheme.
Here we have not yet optimized over gate time $\tau$ and segment number $n_{\mathrm{seg}}$, but as we can see, a high gate fidelity of 99.9\% can already be achieved. In Fig.~\ref{fig:gate} we further show the optimized Rabi frequency $\Omega(t)$ at the detuning $\mu=2\pi\times7.3124\,$MHz [labelled by a black square in Fig.~\ref{fig:scan_fidelity_micro}(b)]. Note that we have $|\Omega(t)|<\mu$ so that strong excitation of the phonon modes can be avoided \cite{PhysRevA.97.062325}.

\begin{figure}[!tbp]
  \centering
  \includegraphics[width=0.7\linewidth]{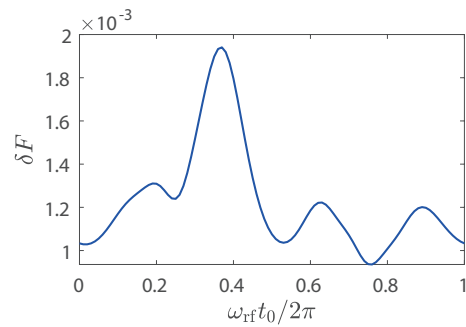}\\
  \caption{Gate infidelity $\delta F$ versus the starting point $t_0$ of the gate during an RF period, using the optimized gate parameters in Fig.~\ref{fig:gate}.}\label{fig:initial_phase}
\end{figure}

In Fig.~\ref{fig:scan_fidelity_micro} we assume that the gate starts right at the beginning of an RF period [$t=0$ in Eq.~(\ref{eq:EOM2})]. If the laser sequence is not locked to the RF signal, or if the time resolution is not high enough, the starting point of the gate can locate anywhere in an RF period, which can be a source of error. In Fig.~\ref{fig:initial_phase} we plot the gate infidelity versus the initial time $t_0$ using the optimized gate parameters in Fig.~\ref{fig:gate}. As expected, the gate infidelity varies with $t_0$ and returns to the original value after one RF period. The variation is not significant and on average we still have a fidelity of $99.88\%$.

In Fig.~\ref{fig:scan_fidelity_micro}(b) we also show a comparison for different truncations of the micromotion. The blue solid curve is computed for a high-order expansion (truncated at $L=5$ for the equilibrium trajectories and $n_{\mathrm{cut}}=5$ for the high-frequency modulation of the normal modes), while the red dashed curve is to apply the same optimized gate sequence on a hypothetical ion crystal truncated at $L=1$ and $n_{\mathrm{cut}}=1$. The high-order calculation is about 50 times slower than the lower-order one, with small but noticeable difference in the gate fidelity. We expect the difference to increase for larger crystals and larger micromotion, for which the high-order computation will be necessary. On the other hand, if we apply the same gate sequence on a hypothetical ion crystal with $L=0$ and $n_{\mathrm{cut}}=0$, that is, without considering micromotion, then the gate infidelity can differ by two orders of magnitude, as shown by the upper green curve. This clearly reveals the nonnegligible effects of micromotion on the gate performance.

As we can see from Fig.~\ref{fig:equilibrium}, in this example, the micromotion along the $z$ direction is much smaller than those in other directions, so we could have gotten much faster convergence had we applied the laser beams in the $z$ direction. This simplification can be used for practical gate design, while here we just want large micromotion to demonstrate our algorithm. Note that the time cost of the algorithm increases linearly with the number of ions (number of normal modes) and linearly with the number of segments (because the bottleneck of the algorithm is the $p=q$ double integral of the two-qubit rotation angle). Also, the scan of the gate parameters can be performed in parallel, and we can first perform a low-order calculation to help locate the parameter regions with potential high gate fidelity. Therefore even though the design of the gate with micromotion is much slower than that without micromotion, we can still expect it to work for hundreds of ions and hundreds of pulse segments.

\begin{figure}[!tbp]
  \centering
  \includegraphics[width=0.95\linewidth]{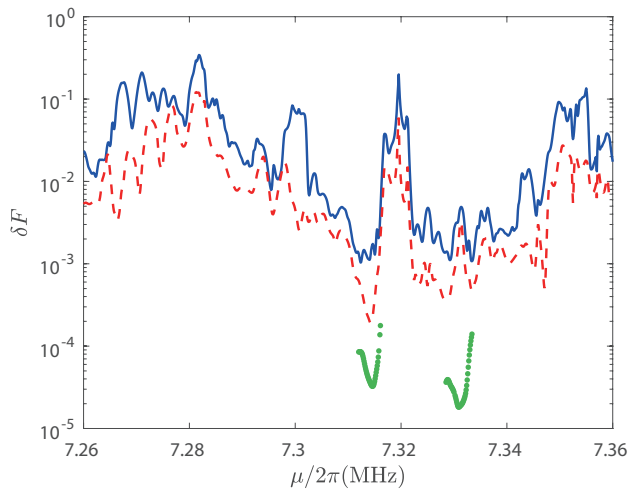}\\
  \caption{Optimal gate infidelity $\delta F$ for a scan of the laser detuning $\mu$ in the same range as Fig.~\ref{fig:scan_fidelity_micro}(b) for $n_{\mathrm{seg}}=15$ (blue solid curve), $n_{\mathrm{seg}}=30$ (red dashed curve) and $n_{\mathrm{seg}}=100$ (green dots). For $n_{\mathrm{seg}}=100$ we only scan around the two regions with the highest fidelity to save the computational resource.}\label{fig:fidelity_nseg}
\end{figure}
\begin{figure}[!tbp]
  \centering
  \includegraphics[width=0.95\linewidth]{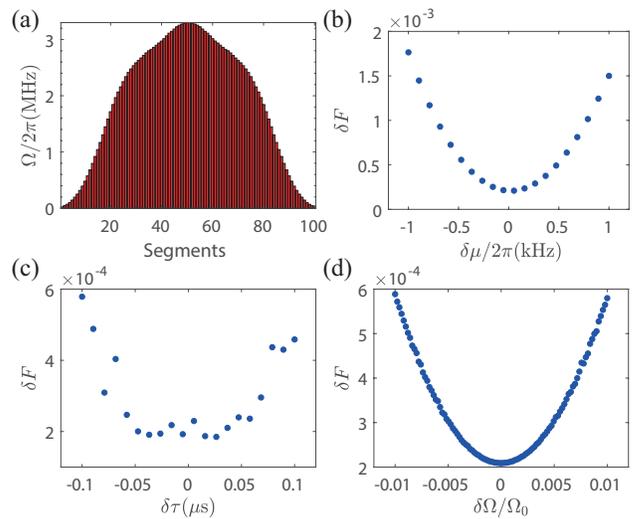}\\
  \caption{(a) Pulse sequence of $n_{\mathrm{seg}}=100$ segments at $\tau=300\,\mu$s and $\mu=2\pi\times 7.3374\,$MHz computed using the method described in Appendix~\ref{app:robustness}. (b) Gate infidelity under the detuning drift $\delta\mu$ of $\pm 2\pi\times 1\,$kHz. (c) Gate infidelity under the gate time drift $\delta\tau$ of $\pm 0.1\,\mu$s. Small fluctuation may be caused by the high-frequency micromotion. (d) Gate infidelity under the fluctuation of $\Omega(t)$ on each segment. Here we assume that each segment is scaled independently by a factor following a Gaussian distribution $N(1,\sigma^2)$ with $\sigma$ up to $1\%$. Each data point is averaged over 1000 random samples and the positive and the negative parts are symmetric.}\label{fig:robustness}
\end{figure}
\section{Discussion and Conclusion}
\label{sec:discussion}
In this work we mainly focus on the effects of micromotion on the gate design and we obtain a theoretical gate fidelity of about $99.9\%$ in Fig.~\ref{fig:scan_fidelity_micro}. This fidelity is mainly limited by the use of small segment number $n_{\mathrm{seg}}=15$ to suppress the residual entanglement to all the $3N=300$ phonon modes, which also exists in earlier works without considering the micromotion \cite{PhysRevA.97.062325}. In Fig.~\ref{fig:fidelity_nseg} we further increase the segment number and observe that the gate infidelity can be reduced to $2\times 10^{-5}$ using $n_{\mathrm{seg}}=100$ segments. To save some computational resources, for $n_{\mathrm{seg}}=100$ we scan the detuning $\mu$ in a more restricted region which is expected to give high gate fidelity from a coarse scan using low-order truncations. With enough control degrees of freedom, the intrinsic gate design error can be made arbitrarily small compared with the other experimental errors.

The robustness against parameter drifts and noise can be studied in the same way as in previous works \cite{PhysRevA.97.062325}. Here we would like to mention that the numerical algorithms we describe in this work are not restricted to the specific method for gate design in Sec.~\ref{sec:gate1d}, and can be easily generalized to take certain robustness conditions against parameter drifts into consideration. For example, in Ref.~\cite{PhysRevLett.120.020501} it has been shown that the robustness of the residual spin-phonon entanglement [Eq.~(\ref{eq:alpha3})] against drift in trap frequency $\omega_k$ can be enforced by requiring the pulse sequence to be symmetric and the time integral of $\alpha_j^k(t)$ over the whole gate sequence to vanish. We can approximately incorporate this criteria into the gate design without increasing the complexity of the algorithm (see details in Appendix~\ref{app:robustness}). Similarly, we can enforce the robustness of the two-qubit rotation angle [Eq.~(\ref{eq:phi_complex3})]. By replacing the optimization conditions with these robustness requirements, we get the pulse sequence presented in Fig.~\ref{fig:robustness}(a) using $n_{\mathrm{seg}}=100$ segments. It can achieve a fidelity above $99.9\%$ for $2\pi\times 0.7\,$kHz drift in laser detuning, $0.1\,\mu$s shift in gate time and $1\%$ random fluctuation of the laser amplitude on each segment. We note that these results are comparable to those in Ref.~\cite{PhysRevA.97.062325} without considering micromotion effects.

To sum up, we have presented a scheme to design entangling gates between two ions in a general ion crystal. The effect of micromotion can be mathematically described as a time-dependent phase of the driving laser, a high-frequency modulation of the laser intensity, and time-dependent mode vectors of ions' collective motional modes. To generalize the gate scheme from a 1D ion chain, first we solve the equilibrium trajectories of the ions for given trapping parameters and then determine the collective normal modes up to arbitrary orders of micromotion. The time integral of the highly oscillatory functions appearing in the gate design is efficiently treated by a series expansion, whose error can be systematically suppressed by including higher and higher order terms. Numerical examples are also provided to show the functioning of our algorithm. Our work lays the foundation for direct quantum computing on a large 2D or 3D ion crystal, and can also help to improve the high-fidelity gate design in 1D ion chain when the transverse micromotion is important.

\begin{acknowledgments}
This work was supported by the National key Research and Development Program of China (2016YFA0301902), Frontier Science Center for Quantum Information of the Ministry of Education of China, and Tsinghua University Initiative Scientific Research Program. Y.-K. W. acknowledges support from Shuimu Tsinghua Scholar Program and International Postdoctoral Exchange Fellowship Program (Talent-Introduction Program).
\end{acknowledgments}

\appendix
\section{Algorithms}
\label{app:algorithm}
In this appendix we briefly summarize the algorithms used in this work.
\subsection{Periodic equilibrium solutions}
To solve the periodic equilibrium solutions satisfying Eq.~(\ref{eq:EOM2}), we take two steps. First we find an approximate solution by introducing a weak damping and then we iteratively improve the accuracy.

\begin{algorithm}[H]
\caption{Approximate periodic equilibrium solutions}
\begin{algorithmic}[1]
\State Choose accuracy target $\epsilon$ and initial damping rate $\gamma$. Start from random initial positions $R_{i\sigma}$.

\State Simulate time-evolution under Eq.~(\ref{eq:EOM2}) plus a damping force $-\gamma\dot{R}_{i\sigma}$ for $N_1$ RF periods. Get final positions $R^{\prime}_{i\sigma}$.

\State If $\|R_{i\sigma}-R^{\prime}_{i\sigma}\|\ge\epsilon$, set $R_{i\sigma}\gets R^{\prime}_{i\sigma}$ and repeat step 2.

\State Simulate time-evolution under Eq.~(\ref{eq:EOM2}) without damping force for $N_2$ RF periods. Get final positions $R^{\prime\prime}_{i\sigma}$.

\State If $\|R_{i\sigma}-R^{\prime\prime}_{i\sigma}\|\ge\epsilon$, reduce $\gamma$ and repeat steps 2-4.
\end{algorithmic}
\end{algorithm}

\begin{algorithm}[H]
\caption{Iteratively approaching periodic equilibrium solutions}
\begin{algorithmic}[1]
\INPUT Approximate periodic solution $R_{i\sigma}(t)$
\State Choose truncation order $L$, accuracy target $\epsilon$ and parameter $\alpha\ge 1$.

\State Compute Fourier series $D_{2n,i\sigma}$ up to orders $\pm L$ using Eq.~(\ref{eq:D_series}) and $G_{2n,ij}$ up to orders $\pm 2L$ using Eq.~(\ref{eq:G_series}).

\State Solve $B_{2n,i\sigma}$ up to orders $\pm L$ using Eq.~(\ref{eq:iteration}).

\State Solve new periodic solution $R_{i\sigma}^\prime(t)$ using Eq.~(\ref{eq:R_fourier}).

\State If $\|R_{i\sigma}(t)-R^{\prime}_{i\sigma}(t)\|\ge\epsilon$, set $R_{i\sigma}(t) \gets R_{i\sigma}^\prime(t)$ and repeat steps 2-4.
\end{algorithmic}
\end{algorithm}
\subsection{Normal modes}
To solve all the normal modes satisfying Eq.~(\ref{eq:coupled_Mathieu}) or more generally Eq.~(\ref{eq:coupled_Mathieu_complete}), we expand the solution into the form of Eq.~(\ref{eq:expansion}).

\begin{algorithm}[H]
\caption{Compute normal modes}
\begin{algorithmic}[1]
\State Choose the accuracy target $\epsilon$.

\State Compute approximate mode frequencies $\beta_k$ and mode vectors $\boldsymbol{C}_0^{(k)}$ using Eq.~(\ref{eq:generalized_eigen}). Compute $\boldsymbol{C}_{\pm 2}^{(k)}$ using Eq.~(\ref{eq:C_approx_2nd}) and $\boldsymbol{C}_{\pm 4}^{(k)}$ using Eq.~(\ref{eq:C_approx_4th}). Use them to construct the approximate vectorial solution $\boldsymbol{C}^{(k)}$ in Eq.~(\ref{eq:mode_matrix_iterative}).

\State For each mode $k$, use $\beta_k$ to compute the matrix in Eq.~(\ref{eq:mode_matrix_iterative}). Compute its eigenvector $\boldsymbol{C}^{(k)\prime}$ with the corresponding eigenvalue $\Delta$ closest to zero, using $\boldsymbol{C}^{(k)}$ as the initial solution.

\State Compute $\beta_k^\prime = \sqrt{\beta_k^2+\Delta}$.

\State If $|\beta_k - \beta_k^\prime|\ge \epsilon$, set $\beta_k\gets \beta_k^\prime$ and $\boldsymbol{C}^{(k)}\gets \boldsymbol{C}^{(k)\prime}$. Repeat steps 3-4.

\State Repeat steps 3-5 for all the modes.
\end{algorithmic}
\end{algorithm}

In the case of degeneracy or near degeneracy of $n$ modes, we can resolve them by solving $n$ eigenvalues closest to zero in step 3. Then we order them as $\Delta_1\le \Delta_2\le \cdots\le \Delta_n$ with corresponding eigenvectors $\boldsymbol{C}^{(k_1)},\,\boldsymbol{C}^{(k_2)},\,\cdots,\,\boldsymbol{C}^{(k_n)}$. For the $i$-th mode, we need to keep using $\Delta_i$ and $\boldsymbol{C}^{(k_i)}$ in the iterations from steps 3-5.

Finally, we can verify that the solved normal modes satisfy the orthonormal condition of Eq.~(\ref{eq:orthonormal}).
\subsection{Gate design}
To design the entangling gate, we need to evaluate integrals in Eq.~(\ref{eq:alpha3}) and Eq.~(\ref{eq:phi_complex3}) with $\chi_j(t)$ given by Eq.~(\ref{eq:chi3}), $\varphi_j(t)$ given by Eq.~(\ref{eq:motional_phi_expansion}), $b_j^k$ replaced by Eq.~(\ref{eq:normal_mode_time_dependent}), and $\Omega_j(t)$ being constant on each segment. Actually, for the specific method of amplitude modulation with piecewise constant $\Omega_j(t)$, we only need to integrate Eq.~(\ref{eq:alpha3}) and Eq.~(\ref{eq:phi_complex3}) for unit $\Omega_j(t)$ on each segment to obtain the row vector $\boldsymbol{A}_j^k$ in Eq.~(\ref{eq:A_mat}) and the matrix $\boldsymbol{\gamma}^\prime(p,q)$ in Eq.~(\ref{eq:gamma_mat}). Then the remaining part of the gate design follows that in Ref.~\cite{PhysRevA.97.062325} as we review in Sec.~\ref{sec:gate1d}.

Now for each element in the row vector $\boldsymbol{A}_j^k$, we have
\begin{equation}
\boldsymbol{A}_j^k (n) = -i\eta_k \sum_\sigma m_{\sigma} \times \mathrm{Eq.~(\ref{eq:50})},
\end{equation}
where $m_\sigma$ is the $\sigma=x,\,y,\,z$ component of the unit vector $\boldsymbol{m}$ of the direction of the Raman laser beams $\boldsymbol{\Delta k}$ and the time integral in Eq.~(\ref{eq:50}) is on the interval $[(n-1)\tau/n_{\textrm{seg}},\, n\tau/n_{\textrm{seg}}]$. Our evaluation of Eq.~(\ref{eq:50}) is based on the following two functions through depth-first search.

\begin{algorithm}[H]
\begin{algorithmic}[1]
\Function{single\_integral}{$t_1$, $t_2$, $\mu$, $\omega$, $\omega_{\mathrm{rf}}$, $L$, $\{\varphi(l)\}$, $\{nk(l)\}$, $k$, $c_0$} \Comment Evaluation of Eq.~(\ref{eq:expand_single})
\State Preset precision $\epsilon$ and cutoff $n_{\mathrm{max}}$
\If{$|c_0|<\epsilon$}
\Return{0}
\EndIf
\If{$k>L$}
\Return{$[c_0\times$prod\_cos($t_1$, $t_2$, $\mu$, $\omega$, $\omega_{\mathrm{rf}}$, $L$, $\{nk(l)\}$, 1)$-c_0^*\times$prod\_cos($t_1$, $t_2$, $\mu$, $\omega$, $\omega_{\mathrm{rf}}$, $L$, $\{nk(l)\}$, 1)$]/2i$}
\Else
\State $nk(k)\gets 0$
\State $v\gets$single\_integral($t_1$, $t_2$, $\mu$, $\omega$, $\omega_{\mathrm{rf}}$, $L$, $\{\varphi(l)\}$, $\{nk(l)\}$, $k+1$, $c_0\times J_0(\varphi(k))$)
\For{$n \gets 1, n_{\mathrm{max}}$}
\State $c\gets 2 i^n J_n(\varphi(k))$
\If {$n > |\varphi(k)|$ and $|c_0\times c|<\epsilon$}
\Break
\EndIf
\State $nk(k)\gets n$
\State $v\gets v+$single\_integral($t_1$, $t_2$, $\mu$, $\omega$, $\omega_{\mathrm{rf}}$, $L$, $\{\varphi(l)\}$, $\{nk(l)\}$, $k+1$, $c_0\times c$)
\EndFor
\State \Return{$v$}
\EndIf
\EndFunction
\end{algorithmic}
\end{algorithm}

\begin{algorithm}[H]
\begin{algorithmic}[1]
\Function{prod\_cos}{$t_1$, $t_2$, $\mu$, $\omega$, $\omega_{\mathrm{rf}}$, $L$, $\{nk(l)\}$, $k$} \Comment Evaluation of Eq.~(\ref{eq:cos_prod})
\If{$k>L$}
\Return{$[e^{i(\mu+\omega)t_2} - e^{i(\mu+\omega)t_1}]/[i(\mu+\omega)]$}
\EndIf
\If{$nk(k)=0$}
\Return{prod\_cos($t_1$, $t_2$, $\mu$, $\omega$, $\omega_{\mathrm{rf}}$, $L$, $\{nk(l)\}$, $k+1$)}
\Else
\State \Return{$\frac{1}{2}$prod\_cos($t_1$, $t_2$, $\mu$, $\omega-k\times nk(k)\times \omega_{\mathrm{rf}}$, $\omega_{\mathrm{rf}}$, $L$, $\{nk(l)\}$, $k+1$)$ + \frac{1}{2}$prod\_cos($t_1$, $t_2$, $\mu$, $\omega+k\times nk(k)\times \omega_{\mathrm{rf}}$, $\omega_{\mathrm{rf}}$, $L$, $\{nk(l)\}$, $k+1$)}
\EndIf
\EndFunction
\end{algorithmic}
\end{algorithm}

The evaluation of Eq.~(\ref{eq:50}) is now given by

\begin{algorithm}[H]
\caption{Evaluation of Eq.~(\ref{eq:50}) for ion $j$, mode $k$ and spatial direction $\sigma$ on the time interval $[t_1,t_2]$}
\begin{algorithmic}[1]
\State Preset precision $\epsilon$ and cutoff $n_{\mathrm{max}}$
\State $v\gets 0$
\State $\{nk(l)\}\gets 0$
\For{$n\gets 0,\pm 1,\cdots,\pm n_{\mathrm{max}}$}
\State $c_0\gets\sqrt{\sum_k |C_{2n,j\sigma}^{(k)}|^2}$
\If{$|c_0|<\epsilon$}
\Break
\EndIf
\State $v\gets v+C_{2n,j\sigma}^{(k)}/c_0\times$single\_integral($t_1$, $t_2$, $\mu$, $\omega_k+n\omega_{\mathrm{rf}}$, $\omega_{\mathrm{rf}}$, $L$, $\{\varphi(l)\}$, $\{nk(l)\}$, 1, $c_0$)
\EndFor
\end{algorithmic}
\end{algorithm}

For the $(p,q)$ element of the matrix $\boldsymbol{\gamma}^\prime$, we have
\begin{equation}
\boldsymbol{\gamma}^\prime(p,q) = \sum_{k\rho\sigma} \eta_k^2 m_\rho m_\sigma \times \mathrm{Im}\left[\mathrm{Eq.~(\ref{eq:53})} + \mathrm{Exch}_{ij} \mathrm{Eq.~(\ref{eq:53})}\right],
\end{equation}
where $\mathrm{Exch}_{ij}$ means exchanging the indices $i$ and $j$ in the following expression. For $p>q$, the double integral in Eq.~(\ref{eq:53}) becomes separable and reduces to
\begin{equation}
\mathrm{Eq.~(\ref{eq:53})}_{ip,jq} = \mathrm{Eq.~(\ref{eq:50})}_{ip} \times \mathrm{Eq.~(\ref{eq:50})}_{jq}^*,
\end{equation}
where the subscript $ip$ means the terms related to the ion $i$ is integrated over the segment $p$ and similarly for the subscript $jq$. The superscript ``$*$'' represents the standard complex conjugate.

To evaluate Eq.~(\ref{eq:53}) for $p=q$, again we use depth-first search by defining the following functions.
\begin{algorithm}[H]
\begin{algorithmic}[1]
\Function{double\_integral}{$t_1$, $t_2$, $\mu$, $\omega_1$, $\omega_2$, $\omega_{\mathrm{rf}}$, $L$, $\{\varphi_1(l)\}$, $\{\varphi_2(l)\}$, $\{nk_1(l)\}$, $\{nk_2(l)\}$, $k$, $c_1$, $c_2$}
\State Preset precision $\epsilon$ and cutoff $n_{\mathrm{max}}$
\If{$|c_1 c_2|<\epsilon$}
\Return{0}
\EndIf
\If{$k>L$}
\Return{
\begin{flushleft}
\setlength{\leftskip}{\algorithmicindent}
$[c_1 c_2 \times$prod\_cos\_plus($t_1$, $t_2$, $\mu$, $\omega_1$, $\omega_2$, $\omega_{\mathrm{rf}}$, $L$, $\{nk_1(l)\}$, $\{nk_2(l)\}$, 1)$- c_1 c_2^*\times$prod\_cos\_minus($t_1$, $t_2$, $\mu$, $\omega_1$, $\omega_2$, $\omega_{\mathrm{rf}}$, $L$, $\{nk_1(l)\}$, $\{nk_2(l)\}$, 1)$- c_1^* c_2\times$prod\_cos\_minus($t_1$, $t_2$, $-\mu$, $\omega_1$, $\omega_2$, $\omega_{\mathrm{rf}}$, $L$, $\{nk_1(l)\}$, $\{nk_2(l)\}$, 1) $+c_1^* c_2^* \times$prod\_cos\_plus($t_1$, $t_2$, $-\mu$, $\omega_1$, $\omega_2$, $\omega_{\mathrm{rf}}$, $L$, $\{nk_1(l)\}$, $\{nk_2(l)\}$, 1)$]/(-4)$
\end{flushleft}}
\Else
\State $v\gets 0$
\For{$n_1 \gets 0, n_{\mathrm{max}}$}
\If {$n_1=0$}
\State $c_1^\prime \gets J_0(\varphi_1(k))$
\Else
\State $c_1^\prime \gets 2i^{n_1}J_{n_1}(\varphi_1(k))$
\EndIf
\If {$n_1 > |\varphi_1(k)|$ and $|c_1 c_2 c_1^\prime|<\epsilon$}
\Break
\EndIf
\State $nk_1(k)\gets n_1$
\For{$n_2 \gets 0, n_{\mathrm{max}}$}
\If {$n_2=0$}
\State $c_2^\prime \gets J_0(\varphi_2(k))$
\Else
\State $c_2^\prime \gets 2i^{n_2}J_{n_2}(\varphi_2(k))$
\EndIf
\If {$n_2 > |\varphi_2(k)|$ and $|c_1 c_2 c_1^\prime c_2^\prime|<\epsilon$}
\Break
\EndIf
\State $nk_2(k)\gets n_2$
\State $v\gets v+$double\_integral($t_1$, $t_2$, $\mu$, $\omega_1$, $\omega_2$, $\omega_{\mathrm{rf}}$, $L$, $\{\varphi_1(l)\}$, $\{\varphi_2(l)\}$, $\{nk_1(l)\}$, $\{nk_2(l)\}$, $k+1$, $c_1 c_1^\prime$, $c_2 c_2^\prime$)
\EndFor
\EndFor
\State \Return{$v$}
\EndIf
\EndFunction
\end{algorithmic}
\end{algorithm}

\begin{algorithm}[H]
\begin{algorithmic}[1]
\Function{prod\_cos\_plus}{$t_1$, $t_2$, $\mu$, $\omega_1$, $\omega_2$, $\omega_{\mathrm{rf}}$, $L$, $\{nk_1(l)\}$, $\{nk_2(l)\}$, $k$} \Comment Evaluation of Eq.~(\ref{eq:theta_integral})
\If{$k>L$}
\Return{integral\_plus($t_1$, $t_2$, $\mu$, $\omega_1$, $\omega_2$)}
\EndIf
\If{$nk_1(k)=0$ and $nk_2(k)=0$}
\State \Return{prod\_cos\_plus($t_1$, $t_2$, $\mu$, $\omega_1$, $\omega_2$, $\omega_{\mathrm{rf}}$, $L$, $\{nk_1(l)\}$, $\{nk_2(l)\}$, $k+1$)}
\EndIf
\If{$nk_1(k)=0$}
\Return{
\begin{flushleft}
\setlength{\leftskip}{\algorithmicindent}
$\frac{1}{2}$prod\_cos\_plus($t_1$, $t_2$, $\mu$, $\omega_1$, $\omega_2-k\times nk_2(k)\times \omega_{\mathrm{rf}}$, $\omega_{\mathrm{rf}}$, $L$, $\{nk_1(l)\}$, $\{nk_2(l)\}$, $k+1$)$ + \frac{1}{2}$prod\_cos\_plus($t_1$, $t_2$, $\mu$, $\omega_1$, $\omega_2+k\times nk_2(k)\times \omega_{\mathrm{rf}}$, $\omega_{\mathrm{rf}}$, $L$, $\{nk_1(l)\}$, $\{nk_2(l)\}$, $k+1$)
\end{flushleft}}
\EndIf
\If{$nk_2(k)=0$}
\Return{
\begin{flushleft}
\setlength{\leftskip}{\algorithmicindent}
$\frac{1}{2}$prod\_cos\_plus($t_1$, $t_2$, $\mu$, $\omega_1-k\times nk_1(k)\times \omega_{\mathrm{rf}}$, $\omega_2$, $\omega_{\mathrm{rf}}$, $L$, $\{nk_1(l)\}$, $\{nk_2(l)\}$, $k+1$)$ + \frac{1}{2}$prod\_cos\_plus($t_1$, $t_2$, $\mu$, $\omega_1+k\times nk_1(k)\times \omega_{\mathrm{rf}}$, $\omega_2$, $\omega_{\mathrm{rf}}$, $L$, $\{nk_1(l)\}$, $\{nk_2(l)\}$, $k+1$)
\end{flushleft}}
\EndIf
\State \Return{
\begin{flushleft}
\setlength{\leftskip}{\algorithmicindent}
$\frac{1}{4}$prod\_cos\_plus($t_1$, $t_2$, $\mu$, $\omega_1-k\times nk_1(k)\times \omega_{\mathrm{rf}}$, $\omega_2-k\times nk_2(k)\times \omega_{\mathrm{rf}}$, $\omega_{\mathrm{rf}}$, $L$, $\{nk_1(l)\}$, $\{nk_2(l)\}$, $k+1$) $+ \frac{1}{4}$prod\_cos\_plus($t_1$, $t_2$, $\mu$, $\omega_1+k\times nk_1(k)\times \omega_{\mathrm{rf}}$, $\omega_2-k\times nk_2(k)\times \omega_{\mathrm{rf}}$, $\omega_{\mathrm{rf}}$, $L$, $\{nk_1(l)\}$, $\{nk_2(l)\}$, $k+1$) $+\frac{1}{4}$prod\_cos\_plus($t_1$, $t_2$, $\mu$, $\omega_1-k\times nk_1(k)\times \omega_{\mathrm{rf}}$, $\omega_2+k\times nk_2(k)\times \omega_{\mathrm{rf}}$, $\omega_{\mathrm{rf}}$, $L$, $\{nk_1(l)\}$, $\{nk_2(l)\}$, $k+1$) $ + \frac{1}{4}$prod\_cos\_plus($t_1$, $t_2$, $\mu$, $\omega_1+k\times nk_1(k)\times \omega_{\mathrm{rf}}$, $\omega_2+k\times nk_2(k)\times \omega_{\mathrm{rf}}$, $\omega_{\mathrm{rf}}$, $L$, $\{nk_1(l)\}$, $\{nk_2(l)\}$, $k+1$)
\end{flushleft}}
\EndFunction
\end{algorithmic}
\end{algorithm}

In the above functions, ``integral\_plus'' is an analytical expression for
\begin{equation}
\int_{t_1}^{t_2} dt \int_{t_1}^t dt^\prime e^{i\mu t} e^{i\mu t^\prime} e^{i\omega_1 t} e^{-i\omega_2 t^\prime}.
\end{equation}
The function ``prod\_cos\_minus'' is defined by making suitable substitution for ``prod\_cos\_plus'' and similarly we define ``integral\_minus'' as an analytical expression for
\begin{equation}
\int_{t_1}^{t_2} dt \int_{t_1}^t dt^\prime e^{i\mu t} e^{-i\mu t^\prime} e^{i\omega_1 t} e^{-i\omega_2 t^\prime}.
\end{equation}
These analytical expressions can be computed easily and here we do not present the detailed expressions for simplicity.

Finally we can evaluate Eq.~(\ref{eq:53}) for $p=q$ as
\begin{algorithm}[H]
\caption{Evaluation of Eq.~(\ref{eq:53}) for ions $i$ and $j$, mode $k$ and spatial directions $\rho$ and $\sigma$ on the time interval $[t_1,t_2]$}
\begin{algorithmic}[1]
\State Preset precision $\epsilon$ and cutoff $n_{\mathrm{max}}$
\State $v\gets 0$
\State $\{nk_1(l)\}\gets 0$, $\{nk_2(l)\}\gets 0$
\For{$n_1 \gets 0,\pm 1,\cdots,\pm n_{\mathrm{max}}$}
\For{$n_2 \gets 0,\pm 1,\cdots,\pm n_{\mathrm{max}}$}
\State $c_1\gets\sqrt{\sum_k |C_{2n_1,i\rho}^{(k)}|^2}$
\State $c_2\gets\sqrt{\sum_k |C_{2n_2,j\sigma}^{(k)}|^2}$
\If{$|c_1 c_2|<\epsilon$}
\Break
\EndIf
\State $v\gets v+C_{2n_1,i\rho}^{(k)}C_{2n_2,j\sigma}^{(k)}/(c_1 c_2)\times$double\_integral($t_1$, $t_2$, $\mu$, $\omega_k+n_1\omega_{\mathrm{rf}}$, $\omega_k+n_2\omega_{\mathrm{rf}}$, $\omega_{\mathrm{rf}}$, $L$, $\{\varphi_1(l)\}$, $\{\varphi_2(l)\}$, $\{nk_1(l)\}$, $\{nk_2(l)\}$, $1$, $c_1$, $c_2$)
\EndFor
\EndFor
\end{algorithmic}
\end{algorithm}

\section{Robustness against trap frequency drift}
\label{app:robustness}
It has been shown in Ref.~\cite{PhysRevLett.120.020501} that robustness of the residual spin-phonon entanglement $\alpha_j^k(\tau)$ against the trap frequency drift $\omega_k$ can be achieved by requiring $\int_0^\tau \alpha_j^k(t) dt=0$. One can easily check that this condition still holds with the existence of micromotion for Eq.~(\ref{eq:alpha3}). Similarly one can check that the robustness of the two-qubit phase $\Theta_{ij}(\tau)$ is ensured by taking the real part rather than the imaginary part in Eq.~(\ref{eq:phi_complex3}) (this comes from the partial derivative with respect to $\omega_k$ which gives us a factor of $i$) and then requiring the time integral to vanish. Evaluating these integrals are again numerically intense. Fortunately, for the numerical algorithms described in the main text and summarized in Appendix~\ref{app:algorithm}, we already get the desired integral on each segment. From these results, not only can we obtain $\alpha_j^k(\tau)$ and $\Theta_{ij}(\tau)$ but also their values at the time points $n\tau/n_{\mathrm{seg}}$ ($n=0,\,1,\,\cdots,\,n_{\mathrm{seg}}$) (for $\Theta_{ij}(\tau)$ we need to keep the complex values before taking the imaginary part), from which we can approximate the their time integral.

Specifically, after computing the row vector
\begin{equation}
\boldsymbol{A}_j^k\equiv [A_j^k(1),\,A_j^k(2),\,\cdots,\,A_j^k(n_{\mathrm{seg}})]
\end{equation}
as described in Appendix~\ref{app:algorithm}, we can construct a new row vector
\begin{equation}
\boldsymbol{\tilde{A}}_j^k\equiv [n_{\mathrm{seg}}A_j^k(1),\,(n_{\mathrm{seg}}-1)A_j^k(2),\,\cdots,\,A_j^k(n_{\mathrm{seg}})],
\end{equation}
such that $\boldsymbol{\tilde{A}}_j^k\boldsymbol{\Omega}$ is proportional to the approximate time integral of $\alpha_j^k(t)$. Similarly, after evaluating all the elements $\gamma^\prime(p,q)$ of the $\boldsymbol{\gamma}^\prime$ matrix (we denote the matrix before taking the imaginary part by $\boldsymbol{\Gamma}^\prime$ such that $\mathrm{Im}[\boldsymbol{\Gamma}^\prime]=\boldsymbol{\gamma}^\prime$), we can define a new matrix $\boldsymbol{\tilde{\gamma}}^\prime$ whose $(p,\,q)$ element is $(n_{\mathrm{seg}}-p+1)\mathrm{Re}[\Gamma^\prime(p,q)]$ ($1\le q\le p\le n_{\mathrm{seg}}$). Then $\boldsymbol{\Omega}^T \boldsymbol{\tilde{\gamma}}^\prime \boldsymbol{\Omega}$ is proportional to the desired time integral for the robustness of $\Theta_{ij}(\tau)$. For the small segment number considered in this work, we are not able to set all these expressions to zeros. Instead, we replace the optimization condition by minimizing
\begin{equation}
\boldsymbol{\Omega}^T \boldsymbol{\tilde{M}}^\prime \boldsymbol{\Omega} + (\boldsymbol{\Omega}^T \boldsymbol{\tilde{\gamma}}^\prime \boldsymbol{\Omega})^2
\end{equation}
under the constraint $\boldsymbol{\Omega}^T \boldsymbol{\gamma}^\prime \boldsymbol{\Omega}=\pm\pi/4$, where $\boldsymbol{\tilde{M}}^\prime\equiv\sum_{jk}\boldsymbol{\tilde{A}}_j^{k\dag} \boldsymbol{\tilde{A}}_j^{k} (2\bar{n}_k+1)$. We further require the pulse sequence to be symmetric \cite{PhysRevLett.120.020501}, namely $\Omega(n)=\Omega(n_{\mathrm{seg}}-n+1)$. This gives us the pulse sequence presented in Fig.~\ref{fig:robustness} in the main text. Note that now the cost function is nonlinear and we can no longer solve it as an eigenvalue problem. However, this cost function is still simple enough with the Jacobian and the Hessian matrix computable, thus the time cost to find an optimal gate design is still negligible compared with the cost for numerical integration.

\bibliographystyle{apsrev4-1}
%\bibliography{reference,D:/Documents/AMO/note/wuyukai}

\end{document}